\documentclass[journal=jacsat,manuscript=article]{achemso}

\usepackage[version=3]{mhchem} 
\usepackage{amsmath}
\usepackage{amssymb}
\usepackage{siunitx}
\usepackage{color}
\usepackage{graphicx}
\usepackage{bm}
\usepackage{color}
\usepackage[b]{esvect} 
\usepackage{mathtools}
\usepackage{braket}
\usepackage{amsfonts}
\usepackage{textcomp} 
\usepackage{microtype} 

\usepackage{hyperref}
\hypersetup{colorlinks=true}
\usepackage[all]{hypcap} 

\newcommand{\mr}[1]{\mathrm{#1}}

\newcommand{\be}{\begin{equation}}
\newcommand{\ee}{\end{equation}}

\newcommand{\kb}{k_{\mr{B}}}

\newcommand{\ec}{E_{\mr{c}}}

\newcommand{\tk}{T_{\mr{K}}}

\newcommand{\vg}{V_{\mr{g}}}

\newcommand{\QD}{quantum dot }

\author{Bivas Dutta}
\author{Danial Majidi}
\author{Alvaro Garcia Corral}
\affiliation{Univ. Grenoble Alpes, CNRS, Grenoble INP*, Institut N\'eel, 38000 Grenoble, France}
\author{Paolo A. Erdman}
\affiliation{NEST, Scuola Normale Superiore and Istituto Nanoscienze-CNR, 56127 Pisa, Italy }
\author{Serge Florens}
\affiliation{Univ. Grenoble Alpes, CNRS, Grenoble INP*, Institut N\'eel, 38000 Grenoble, France}
\author{Theo A. Costi}
\affiliation{Peter Gr\"unberg Institut, Forschungszentrum J\"ulich, 52425 J\"ulich, Germany}
\author{Herv\'{e} Courtois}
\affiliation{Univ. Grenoble Alpes, CNRS, Grenoble INP*, Institut N\'eel, 38000 Grenoble, France}
\author{Clemens B. Winkelmann}
\affiliation{Univ. Grenoble Alpes, CNRS, Grenoble INP*, Institut N\'eel, 38000 Grenoble, France}
\email{clemens.winkelmann@neel.cnrs.fr}

\title{Direct measurement of the Seebeck coefficient in a Kondo-correlated
single-quantum-dot transistor}


\begin{document}
\begin{abstract}
We report on the first measurement of the Seebeck coefficient in a
tunnel-contacted and gate-tunable individual single-quantum dot
junction in the Kondo regime, fabricated using the electromigration
technique. This fundamental thermoelectric parameter is obtained by
directly monitoring the magnitude of the voltage induced in response to a
temperature difference across the junction, while keeping a zero
net tunneling current through the device. In contrast to bulk materials and
single molecules probed in a scanning tunneling microscopy (STM)
configuration, investigating the thermopower in nanoscale electronic transistors
benefits from the electric tunability to showcase prominent quantum effects.
Here, striking sign changes of the Seebeck coefficient are induced by varying
the temperature, depending on the spin configuration in the quantum dot. The
comparison with Numerical Renormalization Group (NRG) calculations demonstrate
that the tunneling density of states is generically asymmetric around the Fermi
level in the leads, both in the cotunneling and Kondo regimes.
\end{abstract}


Exploring charge and heat transport at the level of single atoms or molecules in
contact with voltage and temperature biased reservoirs constitutes the most
fundamental probe of energy transfer at the nanoscale~\cite{Bergfield2013,ThossReview}.
While purely electrical conductance measurements in various quantum dot junctions 
are by now well-established, both experimentally and
theoretically~\cite{PustilnikReview,GrobisReview,BullaReview},
probing electrical and thermal current in fully-controlled nanostructures under 
temperature gradients still constitutes a great experimental challenge. 
The two central thermoelectric quantities are the thermal conductance and 
the thermopower (also known as the Seebeck coefficient).
These relate respectively to the heat current and the voltage resulting from a
thermal imbalance in reservoirs tunnel-coupled through a nano-object, under
the condition of zero net electrical current. Both quantities have been investigated 
at the nanoscale in metallic tunnel
contacts~\cite{Schwab2000,Meschke2006,Jezouin2013,dutta2017thermal} and in
single molecules probed by an STM
tip~\cite{Reddy,widawsky2011,Evangeli,cui2018}. Gate-tunable thermoelectric
experiments, allowing to assess and control the electronic structure of
individual quantum dots, have been conducted so far essentially using
semiconducting structures
\cite{staring1993coulomb,scheibner2005thermopower,svensson2013nonlinear,josefsson2018quantum}.
Conversely, only very few studies in a molecular or nanoparticle transistor
geometry have been performed~\cite{kim2014}, and only with limited gate
coupling.

The rise in nanofabrication techniques has allowed connecting single quantum
dots, small enough to display experimentally reachable level quantization, such
as provided by electrostatically defined regions in 2D electron gases, carbon
nanotubes, single molecules and nanoparticles. This progress has led in recent
years to quantitative understanding of electronic correlations at the
nanoscale~\cite{glazman1988resonant,goldhaber1998kondo,cronenwett1998tunable,
pustilnik2001kondo,sasaki2000kondo,yu2004kondo,roch2008quantum,iftikhar2018tunable,frisenda2015kondo}.
Due to the universal and robust nature of Coulomb blockade and Kondo effects in
single quantum dot electronic junctions, the full characterization of thermoelectric
properties of quantum dots still constitutes a milestone in the field of nanoscale charge
and heat transfer, which delineates the central investigation in this Letter. In
particular, the Kondo effect is a paradigmatic many-body effect of electrons in
bulk metals with magnetic dopants~\cite{kondo1964resistance}, also taking place in 
nanostructures in the regime of Coulomb blockade of the charge with an unpaired magnetic
moment. Driven by the magnetic exchange interaction between the localized
electronic orbital and the conduction electrons near the Fermi level $E_F$, a
hybrid tunneling resonance of width $k_BT_K$ develops at low temperature in the
spectral function near $E_F$, due to the entanglement of conduction electrons to
the quantum dot electronic degrees of freedom, below a characteristic Kondo temperature
$T_K$.

\begin{figure}[t]
\centering
\includegraphics[width=5in]{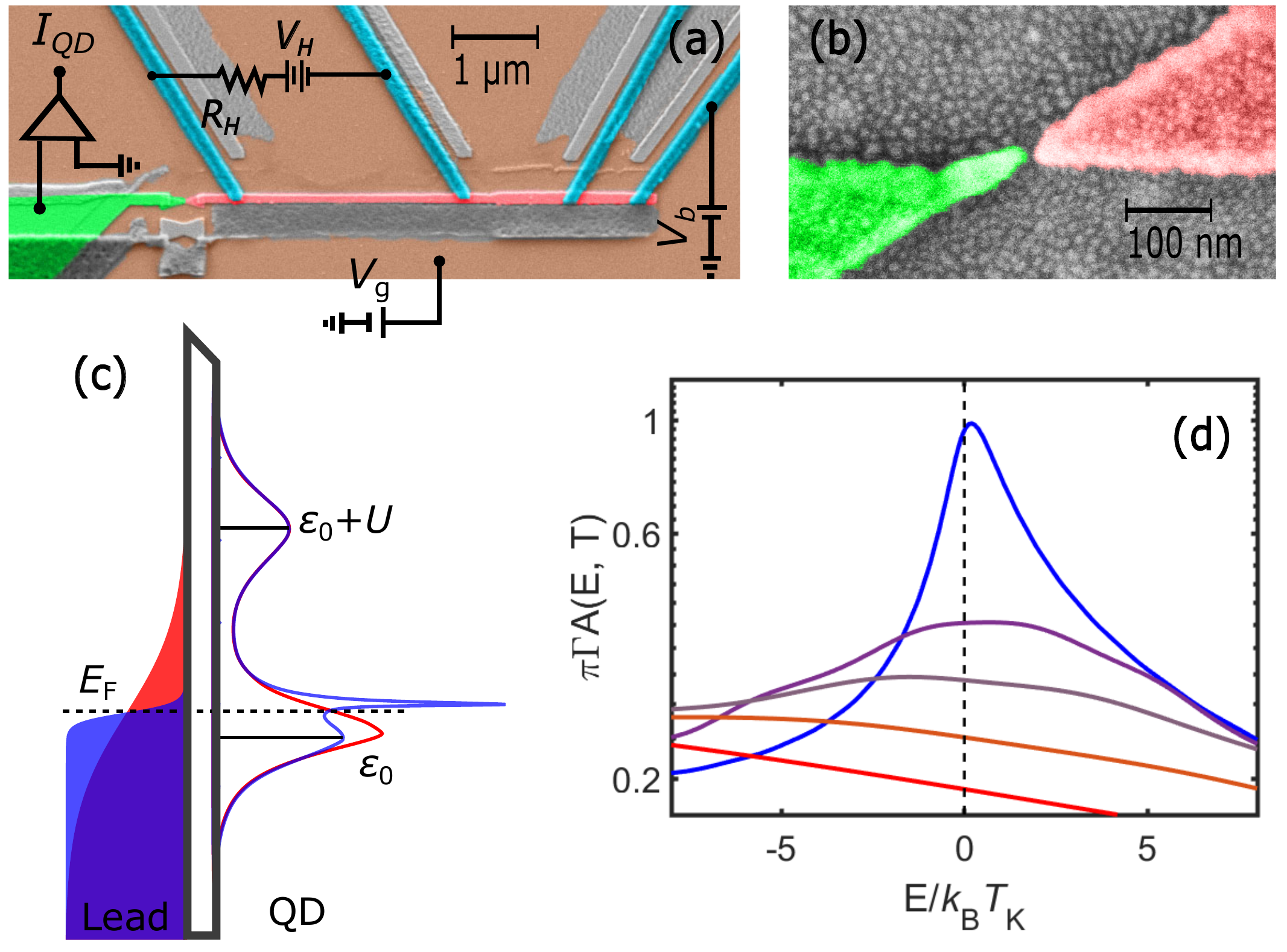}
\caption{ (a) False color scanning electron micrograph of the device, displaying
the drain (green) and source (red) contacts to the quantum dot. Four
superconducting aluminum leads (cyan) are connected to the source, for heating
and biasing the junction. (b) Zoom on an electromigrated quantum dot junction between the
source and the drain. (c) Sketch of the spectral function of the quantum dot (right) 
induced by the coupling to the lead (left), both at high (red) and low (blue) 
temperatures. The Kondo effect arises as a sharp resonance near yet not exactly at
$E_F$ at low temperatures. (d) Numerical renormalization group (NRG) calculation
on the single level Anderson model, with on-site Coulomb interaction $U$ and
level position $\varepsilon_0$.
Here is shown the junction spectral function $A(E)$ at different temperatures $T/T_K=0.01, 2.8, 
5, 10, 20$ (from blue to red), for an asymmetric impurity level and a fixed Kondo 
temperature $T_K$, showing the spectral offset and asymmetry of the 
Kondo resonance.}
\label{fig1}
\end{figure}

Local electrical gating can shift in energy the localized orbital and thus
break particle-hole symmetry. This drives strong thermoelectric 
effects in the tunneling current, associated to spectral asymmetries in the tunneling spectrum.
However, in the standard picture of the Kondo resonance, it is schematically
assumed that the Kondo peak is pinned exactly at the Fermi level, independently
on the depth $\varepsilon_0$ of the localized state. While this picture is
approximately true and amply sufficient to understand roughly the temperature
dependence of the linear conductance in the Kondo regime, it is in fact totally
inadequate for describing the thermopower of quantum dots
\cite{costi2010thermoelectric}.
For gate voltages close to the middle of odd charge Coulomb diamonds, the Kondo resonance 
peak energy differs very little from $E_F$ (by much less than $k_BT_K$), due to
nearly complete realization of particle-hole symmetry. In this regime, thermal transport also
nearly vanishes due to compensating contributions from electron and hole states.
However, this energy shift of the Kondo resonance increases to reach as much as about 
$k_BT_K$ for gate voltages approaching the mixed valence regime in which
the charge on the dot can freely fluctuate, and where one can also anticipate 
enhanced thermoelectric effects (Fig.~\ref{fig1}d) \cite{dong2002effect}.

This asymmetry of the Kondo resonance about the Fermi level, along with its
strong temperature dependence, are crucial for understanding the low temperature
thermopower of Kondo-correlated quantum
dots\cite{costi2010thermoelectric,costi1994transport}. Although well established
by theory \cite{hewson1997kondo}, these properties have not been directly
observed by experiments to date. This is mainly because parasitic voltage
offsets are unavoidable in low temperature transport experiments, due to the
signal amplification chain or thermoelectricity in the wiring, rendering the
precise determination of the Fermi level $E_F$, and thus the relative position of the Kondo peak
with respect to $E_F$, difficult. The situation is different when, in addition
to a voltage bias, a temperature bias $\Delta T$ can be applied across a
junction hosting the Kondo resonance, leading to thermoelectric effects.
Experimentally, the Seebeck coefficient (or thermopower) is defined as
$S=-V_{Th}/\Delta T $ in the linear regime, with $V_{Th}$ the thermo-voltage
established at zero DC current flowing. While the low-temperature linear
conductance probes the amplitude of the junction spectral function $A(E)$ at
$E_F$, the low temperature thermopower is related to the spectral function derivative, 
$S\propto dA/dE\mid_{E_F}$.
More generally, a non-zero Seebeck coefficient in the Kondo state implies that the 
Kondo resonance must be asymmetric about the Fermi level within a temperature window $\pm k_BT$.
Yet, thermoelectric measurements in the presence of Kondo correlations have
remained rare to date\cite{thierschmann2014thesis} and have either focused on
the mixed valence regime\cite{scheibner2005thermopower} or on measurements of
the thermocurrent rather than Seebeck
coefficient\cite{svilans2018thermoelectric}.

Here, we report on a direct measurement of the Seebeck coefficient from the
Coulomb blockade to the Kondo regimes, using combined transport and thermopower
measurements in a single quantum dot junction. From the variations of the
thermopower with level depth at different temperatures, we experimentally verify
two hallmarks of Kondo correlations in thermal transport. First, we report on a
Seebeck signal that is breaking the $1e$-periodicity with respect to the quantum dot charge 
state, which gives strong indication for single-spin induced effects on thermoelectric
properties. Second, we find sign changes in the thermopower upon increasing 
temperature for fixed gate voltages in the Kondo-dominated odd-charge diamonds, 
while no such sign change is observed in the non-Kondo even-charge Coulomb diamonds 
(for fixed gate voltage).
The former reflects the intricate spectral weight rearrangement of the asymmetric 
Kondo resonance from low to high energies as the Kondo peak is destroyed upon
increasing temperature (see Fig.~\ref{fig1}, as well as Fig. S6 in the 
Supporting Information). 
These observations are found in good agreement with
predictions from NRG calculations on the Anderson model described in 
Ref.~\cite{costi2010thermoelectric} and further developed in this work.

Our junctions are realized using the electromigration technique, which has been
successfully applied for studying the Kondo effect in a variety of single
quantum dot systems, such as single molecules and metallic
nanoparticles\cite{roch2008quantum,park2002coulomb,liang2002kondo}. Using
electron-beam lithography and a three-angle shadow evaporation we fabricate
devices such as pictured in Fig.~\ref{fig1}a, on top of a local back gate. After
lift-off, inspection and thus exposure to air, we again evaporate a 1 - 1.5 nm
gold layer over the entire sample surface. Due to its extreme thinness, this
layer segregates into a discontinuous film of Au
nanoparticles\cite{bolotin2004metal}. After cooling to 4.2 K, we form a
nm-scale gap in the platinum constriction visible in Fig.~\ref{fig1}b by controlled
electromigration. Devices displaying reproducible gate-dependent conductance
features are then investigated at temperatures down to 60 mK in a thoroughly
filtered dilution cryostat. The transport properties are determined by measuring
the junction current $I_{QD}$ as a function of the bias voltage $V_b$ and a gate voltage
$V_g$, applied from a local back gate. One lead of the quantum dot junction, defined as
the drain in what follows, rapidly widens away from the electromigration
constriction, allowing for efficient heat draining on that side. In order to
allow the application of a controlled temperature gradient, a normal metallic
wire of length 5 $\mu$m provides the other contact to the quantum dot junction, called the
source. The source side of the junction lead displays four high-transparency
superconducting aluminum contacts. These allow for electrically connecting while
thermally isolating the source at low enough temperatures. Further, we can heat the source electrons by
applying a current between two such leads. In principle, the superconducting
transport properties between two nearby leads across the source can also be used
for local electron thermometry, but due to one missing contact, this was not available in this experiment.

\begin{figure}[H]
\centering
\includegraphics[width=5in]{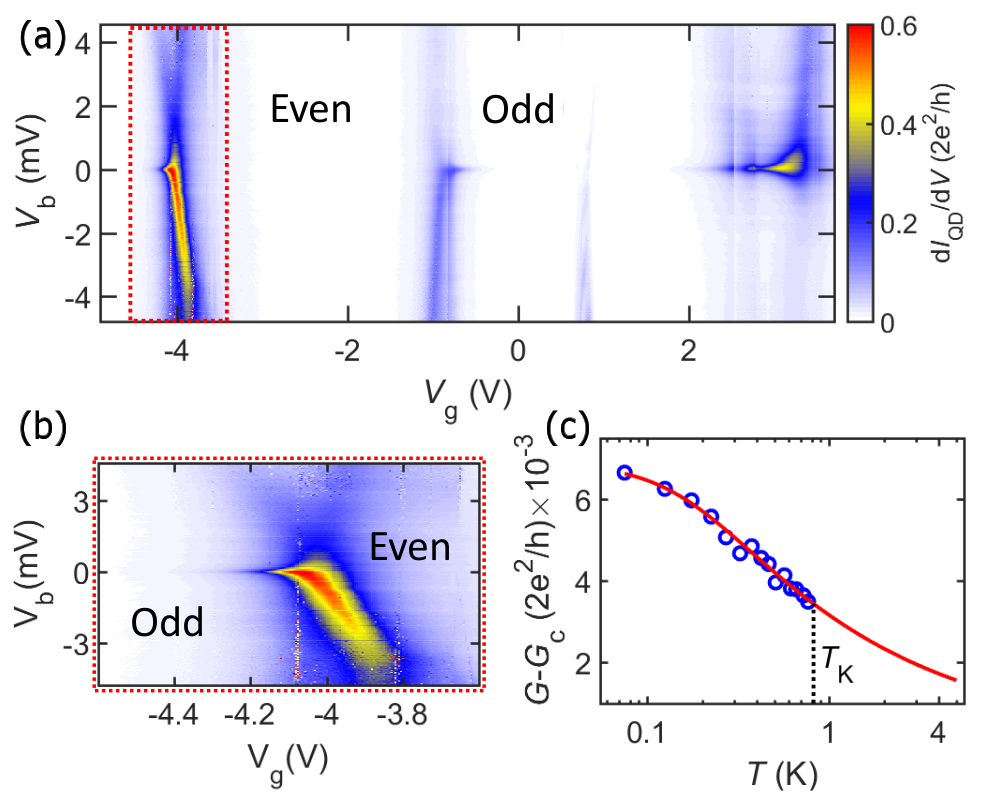}
\caption{ (a) Differential conductance map of the device, measured at base
temperature and without applying a thermal gradient. Three high conductance
degeneracy points separate Coulomb blockaded regions. Every other Coulomb
diamond displays a zero-bias resonance, with decreasing intensity when moving
away from the degeneracy point. From this, the parity of the electron occupation
number can be deduced. Note that the conductance map displays the signature of
another quantum dot connected in parallel to the main device, visible near $V_g=0.7$ V
(see discussion in Sec. III of the Supporting Information). (b) Zoom on the Kondo ridge near
$V_g=-4.2$ V. (c) Temperature dependence of the linear conductance
$G$ (minus a constant background value $G_c\simeq0.004(2e^2/h)$) on the Kondo ridge
at $V_g=-0.295$ V. The line is a fit using a frequently used
phenomenological expression,\cite{goldhaber1998kondo} matching well NRG
calculations. At this gate voltage, $T_K=820$ mK is
defined as the temperature at which the conductance peak height is equal to half
its zero-temperature value.}
\label{fig2}
\end{figure}

Figure~\ref{fig2}a,b shows the differential conductance $G=dI_{QD}/dV_b$ map of the device, as a
function of bias and gate voltage. Four Coulomb diamonds, separated by the
degeneracy points of the quantum dot charge states, can be seen and point to a dot
charging energy $U \approx 58$ meV (in notation of the Anderson impurity model
introduced below). In the device studied here, a second quantum dot appears as a 
faint conductance feature near $V_g=0.8$~V seen in the global transport map of
Fig.~\ref{fig2}a. It has very different transport characteristics and is discussed in
more detail in Sec. III of the Supporting Information. Notably, the thermopower signal
associated to the latter appeared only in a very small gate voltage window, well
separated from that of the more strongly coupled quantum dot. 
The gate voltage region above $V_g =3$ V was subject to electrostatic switches, 
which did not allow accessing quantitatively the full amplitude of the device response 
in this region. In two non-adjacent Coulomb diamonds of the main device ($V_g<-4.0$ V 
and 3.5 V $>V_g> -0.9$ V), a transport resonance near zero bias is observed near
the degeneracy points. This points to a Kondo resonance based on the degeneracy of the
electronic spin$-1/2$ doublet in oddly occupied charge states. From the
temperature dependence of the resonance amplitude (Figure~\ref{fig2}c) we can estimate
the value of $T_K$, which decreases with $\varepsilon_0$ moving towards $-U/2$,
that is, for $V_g$ approaching the center of the odd Coulomb diamond
\cite{goldhaber1998kondo}. 

The peak conductance of the
Kondo resonance saturates at values $< 0.012\times(2e^2/h)$ in the low
temperature limit, from which we can infer that the quantum dot is rather asymmetrically
tunnel coupled \cite{simmel1999anomalous}. This asymmetry simplifies the theoretical description,
as Kondo correlations can be considered as occurring in equilibrium with
the more strongly coupled lead, the other lead acting only as weak probe. In this
study, this strongly coupled lead will thus also serve as the only reference for
the Fermi level, near which the Kondo resonance develops. The tunnel coupling on
the strongly coupled side, $\Gamma \approx 2.6$ meV, can be determined from the
widths of the Coulomb diamond edges (see~\cite{aligia2015impact} for details
pertaining to effects related to the charge parity on the quantum dot, that we
have taken into account). As opposed to most experiments based on semiconducting systems, 
neither the quantum dot nor the tunnel barriers are
electrostatically defined here. Thus $\Gamma$ is essentially independent on the
gate voltage here, which simplifies the theoretical comparison. 

\begin{figure}[H]
\centering
\includegraphics[width=5in]{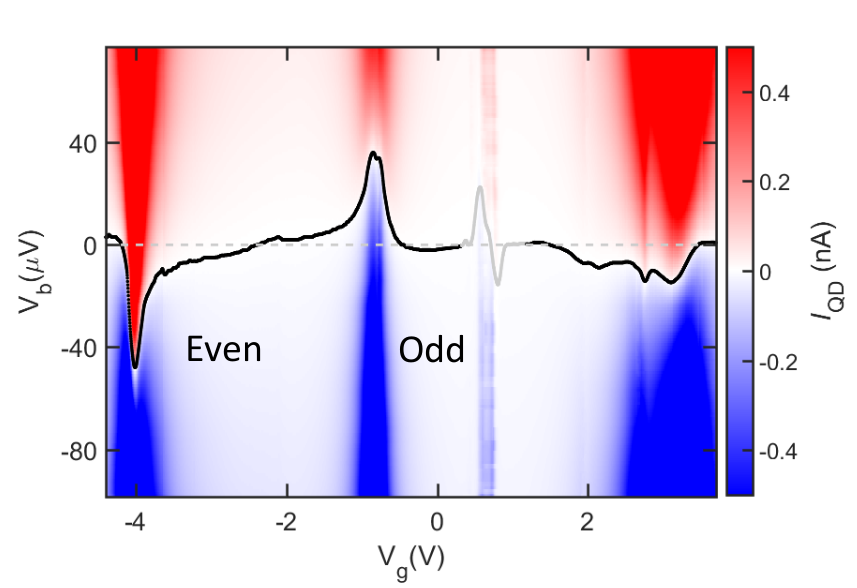}
\caption{Current map for small applied biases, in the presence of a temperature gradient 
at intermediate temperature $T_\mr{mid}\simeq 1$ K. The black line follows the
points of vanishing current; it is thus equal to $-V_{Th}$. The thermoresponse
at about $V_g = 0.7$ V associated to the second, weakly coupled quantum dot, is greyed
out for better readability.}
\label{fig3}
\end{figure}

We now move to the thermoelectric response of the device. We have performed
thermoelectric experiments by providing a constant heating power to the source
island, leading to three device temperatures, labelled
$T_\mr{low}<T_\mr{mid}<T_\mr{high}$. The lowest temperature $T_\mr{low}$ is in the range of a
300 mK, while $T_\mr{high}$ is close to 5 K, and $T_\mr{mid}$ is around 1 K.
Details of the estimation of these temperatures is given in Sec. III of the Supporting 
Information. Measuring the thermopower of a quantum dot junction requires in principle to 
address the open-circuit voltage of a high-impedance device. This is experimentally 
challenging, first because the voltmeter itself may shunt the divergent impedance of the device
and, second, because the equilibration time to reach the true zero-current
state (as required by the definition of the Seebeck coefficient $S$) at such
high impedances can be extremely long. For this reason, several experiments have
preferred focusing on the thermocurrent at zero applied bias rather than on the
Seebeck coefficient, although only the latter has a direct physical
interpretation as a fundamental transport coefficient. In our measurements, we
have used the following, to the best of our knowledge, original protocol: at each 
gate voltage, we sweep the bias
voltage and measure the full $I_{QD}(V_b)$ characteristic (Fig.~\ref{fig3}). 
From thereon, we can define $-V_{Th}$ as the bias voltage at which the current goes through zero,
realizing thus perfect open-circuit conditions. The result is shown as the black
line on the same figure. Strikingly, the thermovoltage changes sign at consecutive
integer charge states, resulting in a $2e$-periodicity of the thermopower
response, that directly follows from the presence of Kondo anomalies in odd
charge diamonds. In more detail, the $2e$-periodicity reflects the fact that the junction 
spectral function has its maximal weight alternating above and below the Fermi level depending on
if the level depth $\varepsilon_0$ of the doublet spin state is either approaching $E_F$ from 
below (in which case the dot transits from single to zero occupancy in the active orbital), or 
$E_F-U$ from above (in which case the highest occupied electronic orbital starts to become singly 
occupied and develops the next Kondo ridge).

While this $2e$-periodic response of thermopower with the quantum dot charge state
is in good agreement with what is expected for the Kondo effect, it is not by
itself a proof thereof. Indeed, in a quantum description of the level
hybridization, the inclusion of the electron spin degree of freedom leads to a
doubling of the spectral function width when the charge states changes 
parity~\cite{aligia2015impact}, breaking thus the $1e$ periodicity naively
expected from a sequential or cotunneling description neglecting the
spin~\cite{beenakker1992theory,turek2002cotunneling}. 
\begin{figure}[H]
\centering
\includegraphics[width=5in]{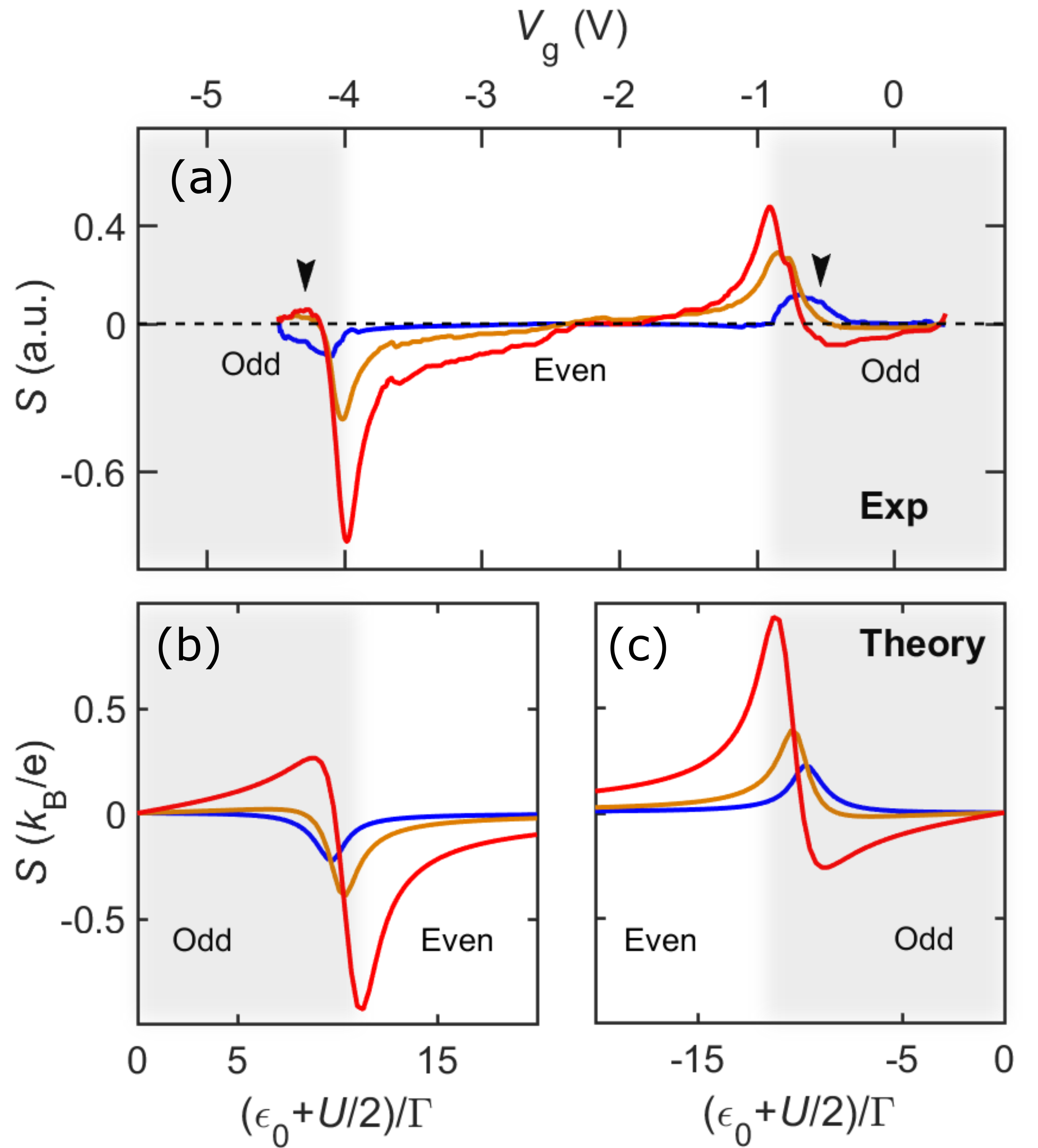}
\caption{(a) Experimental thermopower $S=-V_{Th}/\Delta T$ at
the three experimental device temperatures $T_\mr{low} = 300$ mK
$=0.01\Gamma$ (blue), $T_\mr{mid} = 1$ K $=0.033\Gamma$ (orange) and 
$T_\mr{high} = 4.3$ K = $0.14\Gamma$ (red). 
The arrows highlight the level depths in the Kondo regime
near which the thermopower changes sign at a temperature $T_{1} \approx
\Gamma/(10k_B )$. (b,c) Corresponding NRG calculation using experimental 
parameters $U=58$ meV, $\Gamma=2.6$ meV and for the same set of
temperatures $T/\Gamma$ (with the same color code).
The calculation assumes a single orbital level, predicting
therefore correctly $S=0$ in the center of an oddly occupied Coulomb diamond
($\epsilon_0+U/2=0$). For the sake of comparison with the experimental data, the
calculations at negative $\epsilon_0+U/2$ are placed to the right-hand panel.
Neglecting higher orbital levels in the NRG calculation does not allow to 
map the complete transition region in the center of the even diamond, so that
the theoretical comparison is done using two disjoint panels.}
\label{fig4}
\end{figure}

A much more characteristic signature of the singlet nature of the Kondo state
resides in multiple sign changes of the thermopower as a function of gate
voltage, occuring both in the center of Coulomb-blockaded even and odd
charge states, but also at the onset of the Kondo regime within the odd charge
diamond. This Kondo-related sign change takes place as temperature is increased from 
below to above a characteristic temperature $T_{1}$, which is a weak
function of gate voltage in the Kondo regime (see Fig. S5 of the Supporting Information
and Ref.~\cite{costi2010thermoelectric}). The other Coulomb-related sign changes 
are temperature independent, and occur when the bare quantum dot energy level is 
such that $\varepsilon_0+U/2=0$ (for a single orbital model). In Fig.~\ref{fig4}a we show the gate
traces of the Seebeck coefficient of the same device at different temperatures.
At the lowest temperature $T_\mr{low}$ (such that $k_B T_\mr{low}/\Gamma<0.015$), the 
thermopower inside the Kondo-correlated Coulomb diamonds (for $V_g<-4.1$ V and $V_g>-0.9$ V)
has a markedly different behavior with respect to the higher temperature data,
confirming this sign change.

Our data can be compared with NRG predictions\cite{costi2010thermoelectric} of the 
Seebeck coefficient of a quantum dot with the parameter value $U/\Gamma\simeq22$
taken from the experiment.
The simulations are performed within a two-leads single orbital Anderson impurity model 
with Hamiltonian:
\begin{equation}
\label{model}
H=\sum_{\sigma}\varepsilon_{0}d_{\sigma}^{\dagger}d_{\sigma}^{\phantom{\dagger}} 
+U d_{\uparrow}^{\dagger}d_{\uparrow}^{\phantom{\dagger}} 
d_{\downarrow}^{\dagger}d_{\downarrow}^{\phantom{\dagger}}
+
\sum_{k\alpha\sigma}\epsilon_{k\sigma}
c_{k\alpha\sigma}^{\dagger}c_{k\alpha\sigma}^{\phantom{\dagger}}
+ \sum_{k\alpha\sigma}t_{\alpha}
(c_{k\alpha\sigma}^{\dagger}d_{\sigma}^{\phantom{\dagger}} +H.c.)
\end{equation}
The first term describes the quantum dot level energy $\varepsilon_0$ (measured relative 
to the Fermi level $E_{\rm F}$, which is set to zero in our calculations). The
dot level is controlled in the experiment with the gate voltage $V_g$. The 
second term with charging energy $U$ is the local Coulomb repulsion on the dot. 
The third term describes the Fermi sea in the reservoirs, where $\alpha={L,R}$ labels the 
two contacts, and $\epsilon_{k\sigma}$ is the kinetic energy of the lead electrons.
The last term describes the tunneling of electrons from the leads onto and off the
dot with tunneling amplitudes $t_{\alpha}$. By using even and odd combinations
of lead electron states, the odd channel decouples, resulting in a
single-channel Anderson model with an effective tunnel matrix element $t$ given
by $t^2=t_{L}^2+t_{R}^2$. The hybridization is then characterized by the lead-dot
tunneling rate $\Gamma= 2\pi N_{\rm F}t^2$, with $N_{\rm F}$ the lead electron
density of states at the Fermi level.

The conductance $G(T)$ and thermopower $S(T)$ of the Anderson model~(\ref{model}) can 
be written in terms of the zeroth, $I_0(T),$ and first, $I_1(T)$, moments of the
NRG impurity spectral function $A(E,T)$ within the Fermi temperature window:
\begin{eqnarray}
G(T) &=& G_0 I_{0}(T) \\
S(T) &=& -\frac{1}{|e|T}\frac{I_1(T)}{I_{0}(T)},\\
I_n(T) &=& \int_{-\infty}^{+\infty}-\frac{\partial f(E,T)}{\partial E} E^n\, A(E,T) dE,
\label{Seebeck}
\end{eqnarray}
where $G_0$ is the zero temperature conductance at midvalley and $f(E,T)$ is the
Fermi distribution of the leads at temperature $T$.
The above expression for $S(T)$ implies that a sign change in the thermopower occurs 
due to a crossover between the negative (electron-like) and positive (hole-like) energy 
contributions of the first moment of the spectral function in the Fermi temperature window 
$-k_{B}T < E < +k_{B}T$. At particle-hole symmetric points, such as for perfect integer 
fillings, e.g., exactly in the middle of even or odd Coulomb diamonds, the spectral function 
$A(E,T)$ is symmetric about $E_{F}$, so that $S(T)$ vanishes identically.

The calculated thermopower is plotted in Figs.~\ref{fig4}b and~\ref{fig4}c as a function of the dimensionless 
gate voltage $(\epsilon_0+U/2)/\Gamma$, so that the center of the odd charge
Coulomb diamond at $\epsilon_0=-U/2$ is clearly identified by a trivially vanishing
Seebeck coefficient at that point for any temperature (due to exact particle-hole symmetry). 
Strikingly, the thermopower anomaly seen near the mixed valence regime presents
two distinct regimes: at low temperature, a small thermoelectric signal occurs with a fixed 
sign, while at high temperature a larger signal displays a clear sign change as
a function of gate voltage, defining a crossover temperature $T_{1}$. These predictions compare 
favorably with the experimental data in Fig.~\ref{fig4}a, where the gate-dependent signal shows the same sign 
inversion at temperatures $k_B T_{1} /\Gamma\approx0.1$, depending slightly on
the gate voltage (see Fig. S5 of Supporting Information).
Ultimately, this sign change of the Seebeck coefficient $S$ upon increasing temperature from 
$T<T_{1}$ to $T>T_{1}>T_K$ reflects the spectral weight rearrangement of the asymmetrically located 
Kondo resonance about $E_F$ (see Fig.~\ref{fig1}d, and Sec. IV of the Supporting Information for details).

In conclusion, this work provides a direct measurement of the Seebeck
coefficient for a Kondo-correlated single quantum dot tunnel
coupled to purely thermal-biased reservoirs. In particular, our measurements
bring compelling experimental evidence for a frequently overseen property of the
Kondo effect occurring between a spin-degenerate local level and an electron
reservoir. By measuring the temperature and gate dependence of the Seebeck
coefficient in a single quantum dot junction, we find that it exhibits
characteristic sign changes in the Kondo regime upon increasing temperature,
which reflect the strong temperature dependence of the Kondo peak that is not
exactly pinned at the reservoir Fermi level, as predicted by theory. This work
finally demonstrates that electromigrated single quantum dot junctions can now
be integrated into more complex circuits, including local electronic heaters and
thermometers. This development paves the way for precisely accessing the
thermoelectric figure of merit of individual molecules, which requires measuring
simultaneously the charge and heat conductance as well as the thermopower, for a
large spectrum of molecular devices.


\begin{acknowledgement}

This work has received funding from the European Union's Horizon 2020 research
and innovation programme under the Marie Sk\l{}odowska-Curie grant agreement No
766025. B.D. acknowledges support from the Nanosciences Foundation, under the
auspices of the Universit\'e Grenoble Alpes Foundation. The samples were
realized at the Nanofab platform at Institut N\'eel, with extensive help from T.
Crozes. We are deeply indebted to F. Taddei for theoretical guidance about the
cotunneling analysis of the thermovoltage. Supercomputer support by the John von 
Neumann Institute for Computing (J\"ulich) is gratefully acknowledged. We further 
acknowledge inspiring discussions with N. Roch and J. Pekola.

\end{acknowledgement}

\providecommand{\latin}[1]{#1}
\makeatletter
\providecommand{\doi}
  {\begingroup\let\do\@makeother\dospecials
  \catcode`\{=1 \catcode`\}=2 \doi@aux}
\providecommand{\doi@aux}[1]{\endgroup\texttt{#1}}
\makeatother
\providecommand*\mcitethebibliography{\thebibliography}
\csname @ifundefined\endcsname{endmcitethebibliography}
  {\let\endmcitethebibliography\endthebibliography}{}

\setcounter{figure}{0}
\setcounter{table}{0}
\setcounter{equation}{0}

\global\long\def\theequation{S\arabic{equation}}
\global\long\def\thefigure{S\arabic{figure}}
\renewcommand{\thetable}{S\arabic{table}}

\vspace{1.0cm}
\begin{center}
{\bf \large Supporting information for ``Direct measurement of the Seebeck 
coefficient in a Kondo-correlated single quantum dot transistor''}
\end{center}

\vspace{1.0cm}

{\small
This supporting information discusses the sample fabrication process, the electrical 
characterization of the Kondo-correlated quantum dot, the determination of average 
temperatures of the device under heating conditions, and gives an outline of the theory 
of the thermoelectric properties of Kondo correlated quantum dots, relevant to the 
present paper. Numerical Renormalization Group (NRG) calculations provide also new insights
and detailed analysis on the thermopower of these systems.
}

\vspace{1.0cm}

\section{I - Sample fabrication}

We fabricate the sample on a 2 inch Si $<$100$>$ wafer with a 500 nm thermal
oxide. The first step of the fabrication process is the gate layer.  We use a
metallic plane covered with an oxide layer as the gate of the \QD device. The
reason for choosing this local back-gate is that, using this gate configuration, 
we can easily achieve a very small distance ($<$ 10 nm) between the gate and the
gold nano-particles and thereby achieve a strong gate coupling. We use
laser lithography to pattern the gate structures on top of a cleaned Si wafer
coated with a double layer of photo resists LOR3A/S1805. After development of
the exposed area, we evaporate 3 nm titanium (Ti), 30 nm of gold (Au) and
again 3 nm of Ti. Both Ti layers act as an adhesive for the following layers.
After liftoff and cleaning, the wafer with metallic gate structures is coated
with an approximately 8 nm of $Al_2O_3$ layer using the atomic layer deposition
(ALD) technique.

\begin{figure}[H]
\includegraphics[width=0.55\columnwidth]{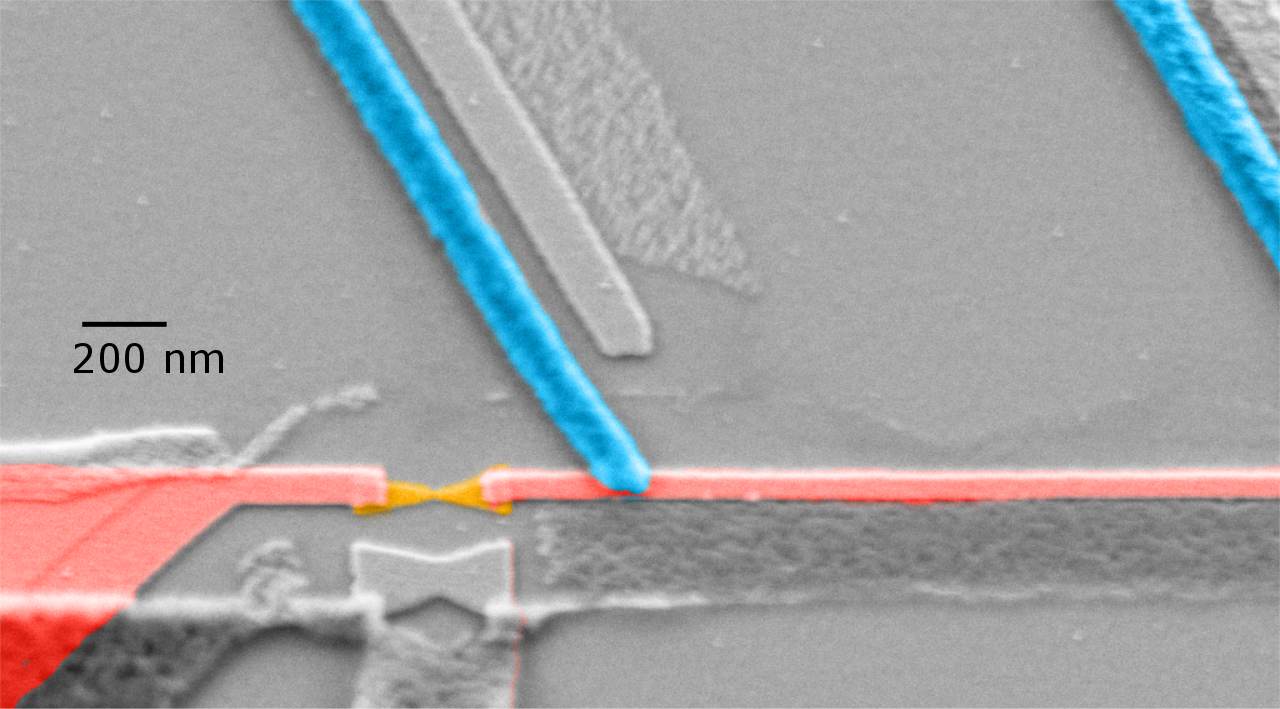}
\caption{False-color scanning electron micrograph of the sample made by three
angle shadow evaporation technique. The colors indicate the three metals:
\textit{bow-tie} shaped $Pt$ constriction is shown as the bowtie in yellow, the Au/Ti-made
\textit{source} and \textit{drain}, connected through the constriction, are shown
in red and the four Al probes (two shown here), connected to the
\textit{source} via transparent contact, are shown in cyan color.}
\label{fig:sample}
\end{figure}

We designed the main parts of the \QD device on top of this gate using
electron-beam lithography. The source, constriction, drain and the four probes
of the device are patterned on top of the processed wafer coated with a double
layer of e-beam resists P(MMA-MAA) 9$\%$ and PMMA 4 $\%$. After development of
the exposed area, we load the sample in an e-beam evaporator with rotatable
sample stage to evaporate metals. First, we deposit 11 nm of platinum (Pt) at
an angle of \ang{-42} w.r.t. the source of the evaporator, this forms the
\textit{bow-tie} shaped constriction of the device, indicated by yellow color in
Fig. \ref{fig:sample}. Then we rotate the sample stage and deposit 25 nm of Au
at an angle of \ang{-22}, which forms the source and drain of the device (red
color in Fig. \ref{fig:sample}), on top of the Pt constriction. At the same
angle, a 3 nm of $Ti$ is then deposited to protect the Au layer from intermixing
with the following aluminum (Al) layer. After that we rotate the sample to
\ang{+20} and deposit 80 nm thick Al contacts, which form the four Al probes on top
of the Au source with a clean interface (cyan color in Fig. \ref{fig:sample}),
making four $S-N$ junctions. After liftoff with acetone, IPA and cleaning with
$O_2$ plasma, the device is ready for the nano-particle deposition.

In order to form the gold nano-particles, we evaporate 1-1.5 nm of Au on top of the 
as-made sample. Due to extreme thinness and to surface tension forces, the evaporated 
Au forms a self-assembled layer nano-particles on top of the sample
\cite{bolotin2004}. An SEM image of such evaporated nano-particles is shown in Fig. 1(b) of 
the main paper. The average size of the gold nano-particles lies in the range of 5-10 nm,
which serves as the \QD in our device. The advantage of this technique of nano-particle
deposition is that, due to high density of the nano-particles the yield of successful
single-\QD device is rather high (about 70 $\%$).

To complete the fabrication process and place a \QD in between the $source$ and
$drain$ lead, we electromigrate the constriction, by passing a current through
it in a controlled manner \cite{ParkEM1999, strachanEM2005, WuFeedbackEM2007}.
As a result, a nano-gap is created between the $source$ and $drain$ leads,
bridged by one or sometimes several of the previously deposited gold nano-particles. 
In order to achieve a strong tunnel coupling between the leads and quantum dot, we 
perform the electromigration inside the cryostat at 4 K and under cryogenic vacuum.

\section{II - Characterization of the quantum dot junction}

\subsection{Quantum dot parameters}

The linear-conductance map of the \QD junction is shown in Fig. 2(a) of the
paper. The parameters of the \QD can be extracted from this conductance map.
The positive and negative slopes of the Coulomb diamond edges for the
Kondo-coupled \QD gives the estimation of the gate coupling factor, also known
as lever-arm, $\alpha \approx$ 0.02 and the capacitive asymmetry between the
source and drain $C_d/C_s = 0.35$. The gate voltage difference between the two
consecutive charge degeneracy points gives the estimate of the charging energy
($\ec$) of the Kondo-coupled \QD as $U = 2\ec$ $\approx$ 58 meV. The relative
contrast of the two Coulomb edges indicates a strong asymmetry in the tunnel
coupling of the \QD to the leads. The total tunnel rate $\Gamma = \Gamma_s+
\Gamma_d$ can be extracted from the width of the Coulomb edges as  $\hbar\Gamma
= 2.6$ meV $\equiv 30$ K, taking into account charge parity effects
on the renormalization of the linewidth~\cite{Saligia2015impact}.

The weakly-coupled dot, which is coupled in the gate voltage range 0.6 V$<$
$V_g$ $<$ 1 V, is characterized in the same way. The slopes of the Coulomb
diamond edges give the estimate of the lever-arm $\alpha \approx$ 0.02. The
charging energy U/2 of the weakly coupled \QD cannot be measured exactly since
we have not observed any other diamond corresponding to this \QD. Therefore only
a lower limit can be given as U/2 $>$ 200 meV.

\subsection{Study of Kondo Effect}

The Kondo spin-singlet formed by the unpaired electron in the \QD and the
conduction electrons with opposite spin in the lead forms below the
characteristic Kondo temperature $T_\mr{K}$\cite{Wilson1975}. We characterize
the Kondo effect in the \QD junction in two different ways, firstly, by
measuring the Kondo-resonance peak as a function of temperature and secondly, by
measuring the splitting of the Kondo-resonance with the application of a
magnetic field \cite{grobis2007kondo}.

The Kondo-resonance is very sensitive to the temperature, and the DC value 
of the conductance at zero-bias reduces with temperature. One can extract
$\tk$ by fitting the temperature dependence of the conductance at the
Kondo-resonance with the empirical formula based on NRG simulations
\cite{CostiNRG1994,Gordon1998}:

\be
G(T)=G_l \left(\frac{T^2}{T_K^2}\left(2^{1/s}-1\right)+1\right)^{-s}\!\!+ G_c,
\label{NRG_TK}
\ee
where $G_l = \frac{2e^2}{h} \frac{4\Gamma_s\Gamma_d}{(\Gamma_s+\Gamma_d)^2}$ is
the saturation value of the conductance at the lowest bath temperature. In the
case of a symmetric coupling of the leads to the \QD, $G_l = 2e^2/h = G_0$, the
quantum of conductance. But for an asymmetric tunnel coupling only a fraction of
the quantum-conductance is achieved in the zero temperature limit, since the
Kondo-resonance develops only with the strongly coupled lead while the other lead
acts as a probe \cite{simmel1999}. $G_c$ is the background conductance due to
the direct tunnelling or conduction through a highly resistive shunt across the
\QD junction. Here, one can set phenomelogically $s = 0.22$ for the spin-1/2 Kondo-effect.
\begin{figure}[H]
	\includegraphics[width=0.9\columnwidth]{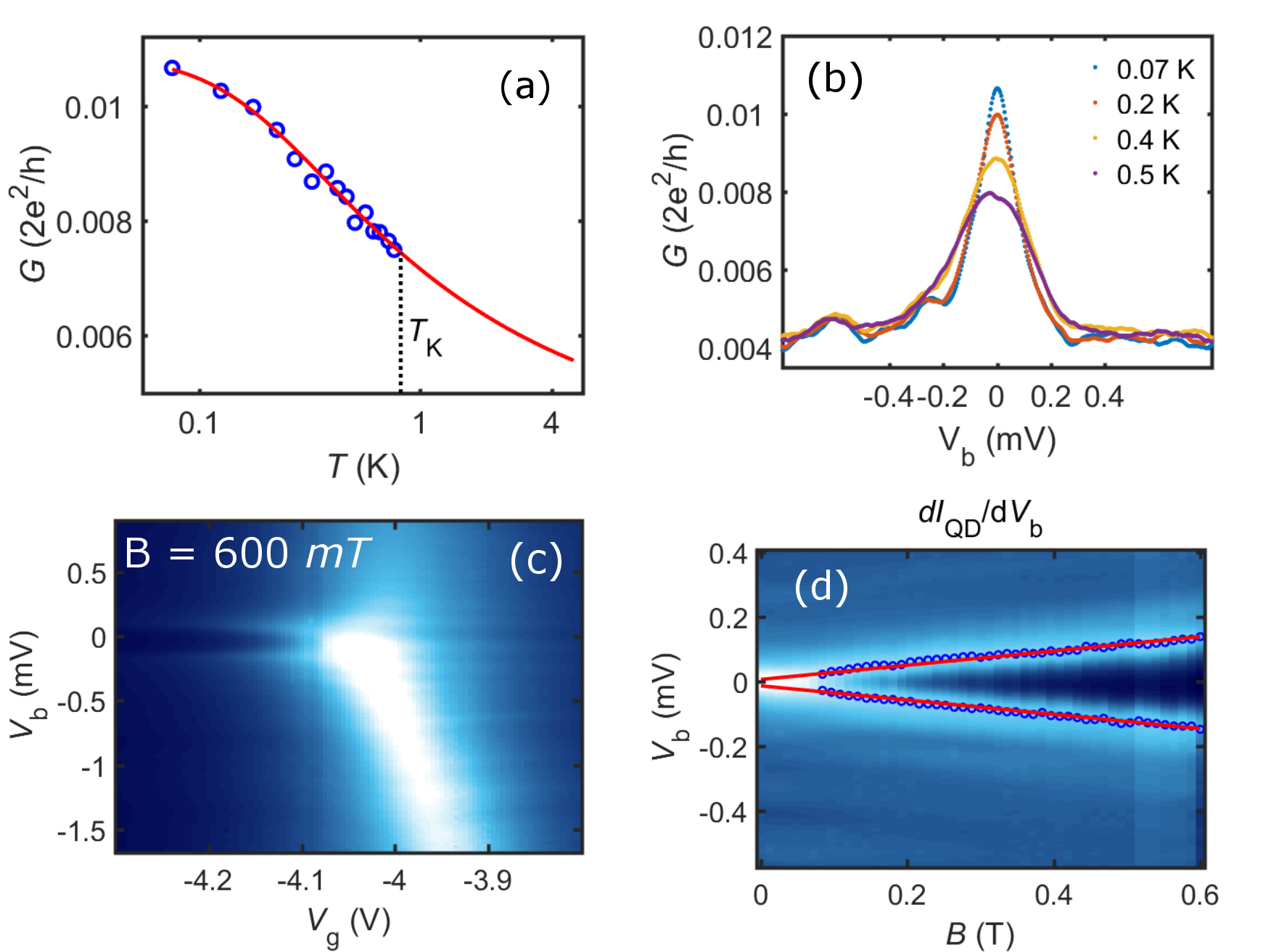}
	\caption{Characterization of the Kondo-effect: (a) The evolution of the 
Kondo-conductance peak with temperature at a fixed gate voltage $\vg = -0.295$ $V$. 
Red curve is the fit of the data with the NRG theory given approximattely by Eq. 
(\ref{NRG_TK}), using $\tk$ and $\Gamma^l/\Gamma^r$ as fitting parameters and a 
constant finite-bias background conductance $G_\mr{c} = 0.004 (2e^2/h)$ (as seen
in panel (b)). 
The extracted Kondo temperature at this gate voltage is $\tk = 0.820$ K. (b) Bias
trace of the conductance at different temperatures. Finite-bias conductance
gives the value of $G_\mr{c}$. (c) Splitting of the Kondo-peak in the presence
of a constant magnetic field B = 600 mT, as a function of gate and bias
voltages. (d) Map of the Kondo-peak conductance with bias voltage and magnetic field. 
The Kondo peak splits at a critical magnetic field $B_c$.} \label{fig:Kondo_chrac}
\end{figure}
The Kondo-temperature is defined as the temperature at which the
Kondo-conductance peak value is reduced to 50 $\%$ of the conductance peak at
the lowest temperature \cite{Gordon1998}. Therefore, using the above formula,
$G(\tk) = G_l/2+G_c$. Fig. \ref{fig:Kondo_chrac} (a) shows the evolution of the
Kondo-conductance peak with temperature at a fixed gate voltage $\vg = -0.295$
V. The red solid-curve shows the fitting of the data with the above formula,
using $\tk$ and $\Gamma_s/\Gamma_d$ as the fitting parameters. From the fit we 
extract the Kondo temperature of the \QD junction at the same gate position as 
$\tk = 0.820$ K and a tunnel coupling asymmetry $\Gamma_s/\Gamma_d = 0.002$.

Another, but less precise way to determine the Kondo temperature is by perturbing 
the Kondo state with the application of a magnetic field. 
A large splitting of the Kondo peak at a fixed magnetic field of $600$ mT is shown 
in Fig.  \ref{fig:Kondo_chrac} (c). The splitting develops beyond at a critical
magnetic field $B_c$ so that $g\mu_BB_c \simeq 0.5\kb\tk$, where $\mu_B$ is
the Bohr magneton and $g$ is the Land\'e g-factor (note that this scale $T_K$
differ by a numerical prefactor from the Kondo scale determined above using the 
temperature dependence of the zero-bias conductance).
We measured the conductance of the \QD junction as a function of bias and 
magnetic field. The conductance of the \QD junction at a constant gate voltage,
with varying magnetic field and bias is plotted in Fig. \ref{fig:Kondo_chrac} (d). The
splitting of the Kondo-resonance is resolved at a magnetic field $B_c \approx$
60 $mT$.
In addition, since splitting of the level is proportional to the applied bias
at large bias, it can be fitted with a linear equation, $eV_\mr{b} = g\mu_BB$. 
The fitted red lines in Fig. \ref{fig:Kondo_chrac} (d) give an estimate of the 
Land\'e-g factor $g = 3.6$, that is somewhat larger than what is usually found in 
gold nanoparticles \cite{bolotin2004}. Using the measured critical field that is 
necessary to split the Kondo resonance, we deduce from this procedure an approximate 
estimate of the Kondo temperature $T_K \simeq 300$ mK, that is of the same
magnitude but somewhat smaller than the Kondo scale extracted from the
temperature dependence.

\section{III - Thermoelectric measurement}

\subsection{Experimental temperatures under heating conditions}

The experimental device temperature and the thermal gradients under which the three
thermopower traces shown in Fig. 4 of the main paper have been obtained
cannot be controlled independently in the experiment, as they depend on the
thermalization process of the device under the applied heating.
The thermal experimental conditions of the three measurements
are summarized in TABLE \ref{table:Tavgs}.
\begin{table}[h]
	\caption[Device temperatures]{Heating conditions and bound estimates
on the device temperatures for the measurements in Fig. 4 of the
main text}
	\centering
	\label{table:Tavgs}
	\begin{tabular}{|c| c| c| c|}
		\hline
		Exp & $T_{cryostat}$ (K) & ${\dot Q^{\phantom{A^A}}}_{H}$ (nW) & $T_{source}$ (K) \\
		\hline
		$T_\mr{low}$ & 0.075  & 0.001  & $\leq$0.52  \\
		\hline
		$T_\mr{mid}$ & 0.075 & 2.7  & $\leq$2.5 \\
		\hline
		$T_\mr{high}$ & 4.2 & 2.7  & $4.2\leq T_{source}\leq$5 \\	
		\hline
	\end{tabular}
\end{table}

There is an evident hierarchy $T_\mr{low}<T_\mr{mid}<T_\mr{high}$ in the set of
operation temperatures. We give here upper bounds on each of these, simply by making a heat-balance between the input
constant Joule heating ${\dot Q}_{H}$ to the source island and the heat leak
associated to electron-phonon coupling ${\dot Q}_{e-ph}$, $${\dot Q}_{H}-{\dot
Q}_{e-ph} = 0,$$ where ${\dot Q}_{e-ph} = \Sigma {\mathcal V} \left(
T_{source}^5 - T_{base}^5\right)$ relates the steady state heat flow to the
interaction volume ${\mathcal V} $ of the source island, the electron-phonon
interaction constant\cite{Echternach1992} in gold $\Sigma=2.4\times10^9$
W.m$^{-3}$.K$^{-5}$ and the two respective temperatures. Such an analysis has
been successfully applied in several previous works (see \cite{DuttaPRL2017} for
example). Because heat can leak out of the source side of the device also by
other channels, mainly by electronic conduction through the leads, the true
device temperature must be substantially below that such estimate, in particular at
temperatures higher than several hundred mK, at which the superconductivity in
the leads is weakened.

At a cryostat temperature of 4.2 K, the electron-phonon coupling (and thermal
conductances of the leads and of the substrate) are large. Thus the thermal
imbalance created across the junction using a heating power of 2.7 nW is barely
sufficient to heat the source electrode to 5 K. At a cryostat temperature of 75
mK, a heating power of 1 pW will lead to $T_{source}=520$ mK if one neglects all
other heat leak channels. At the same cryostat temperature, a heating power of
2.7 nW cannot heat the device above 2.5 K following the above arguments, but the
true device temperature is presumably closer to 1 K. Because
$T_{drain}<T_{source}$, the estimations of the $T_{source}$ are evidently  an
upper  bound to the device average temperature in each case.

In conclusion, $T_\mr{low}$ is of the order of 300 mK, $T_\mr{high}$ is
slightly above 4 K and $T_\mr{mid}$ has an intermediate value, on the order of 1
K. These estimates have been taken for the temperatures used in the Numerical 
Renormalization Group (NRG) calculations, with an hybridization
$\Gamma /k_B\approx$ 30 K. The obtained theoretical predictions for the
thermopower, shown in Fig. 4b)c) of the main manuscript, are in good agreement
with the experimental measurement of Fig. 4a).

\subsection{Thermovoltage signal: Kondo dot vs weak dot}

\begin{figure}[H]
	\includegraphics[width=0.6\columnwidth]{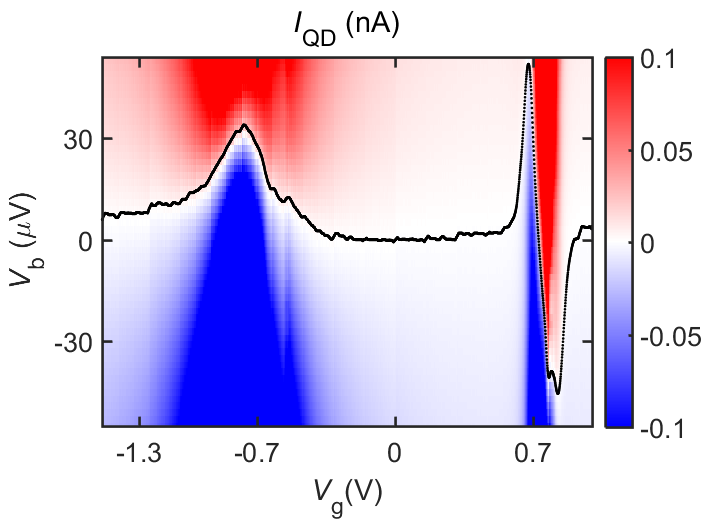}
	\caption{Current map of the \QD junction, hosting a Kondo-coupled dot
and a weakly coupled dot, in the presence of a temperature gradient (device
temperature $T_\mr{mid}$). The black line traces the gate voltage dependence of the
thermovoltage ($-V_{TH}$), determined by the bias voltage for which the current
is vanishing. The distinct features of the thermovoltage line-shape
therefore distinguishes between the Kondo-coupled dot (asymmetric peak with a
shallow zero, on the left side) and weakly coupled dot (symmetric double peak with
a sharp zero, on the right side).}
\label{fig:vth_kondo_weak_dot}
\end{figure}

The measurement of the thermovoltage of the \QD junction shows clearly distinct
features for the Kondo-coupled dot and the weakly coupled dot respectively. The
thermovoltage is determined in the same manner as described in the main paper:
we increase the bias across the \QD junction and measure the current and
therefrom define the thermovoltage $-V_{Th}$ as the voltage bias where the
measured current becomes zero. The current map of the \QD junction with the
black line-trace for the thermovoltage is shown in Fig.
\ref{fig:vth_kondo_weak_dot}.
As it can be seen in the figure, the thermovoltage signal for the Kondo-coupled
dot in the gate voltage range $-1.3V < \vg < 0V$ has a positive sign, while for
the dot that is coupled in the gate voltage range  $0.6V < V_g < 0.8V$, the
thermovoltage signal has both positive and negative sign and passing through
zero at the degeneracy point at $V_g \approx 0.7$ V, which is the expected
thermovoltage-signature of a weakly coupled dot \cite{BeenakkerPRB1992,
TurekPRB2002}.

\section{IV - Numerical Renormalization Group approach to
  thermoelectricity}
\label{sec:NRG}
In the following, we describe the parameter regimes of the model 
used to describe the thermopower of Kondo correlated quantum dots,
outline the transport calculations via the numerical renormalization
group, summarize the temperature and gate voltage dependence of the
thermopower \cite{Costi2010} and provide a more detailed analysis of the temperature
induced sign changes of the thermopower in the Kondo regime. In particular,
we explain the origin of the sign change at $T_{1}$ 
in terms of a temperature dependent spectral
weight shift of the  asymmetrically located Kondo resonance. Finally,
estimates for the gate voltage dependence of $T_1$ for the parameters
of the experiment are given.
\subsection{Model and parameter regimes}
\label{subsec:intro}
In order to describe the thermoelectric transport through a strongly correlated
quantum dot we use a two-lead single-level Anderson impurity model. The
Hamiltonian is given by $H=H_{\rm dot}+H_{\rm leads}+ H_{\rm tunneling}$. The
first term, $H_{\rm
dot}=\sum_{\sigma}\varepsilon_{0}d_{\sigma}^{\dagger}d_{\sigma}
+Un_{d\uparrow}n_{\downarrow}$, describes the dot Hamiltonian, where
$\varepsilon_0$ is the level energy, measured relative to  the Fermi level
$E_{\rm F}$, $n_{d\sigma}$ is the occupation number for spin
$\sigma={\uparrow,\downarrow}$ electrons in the dot level, and $U$ is the local
Coulomb repulsion on the dot. The second term describes the Hamiltonian of the
leads and is given by $H_{\rm
leads}=\sum_{k\alpha\sigma}\epsilon_{k}c_{k\alpha\sigma}^{\dagger}c_{k\alpha\sigma}$,
where $\alpha={L,R}$ labels the two leads, and $\epsilon_{k\sigma}$ is the
kinetic energy of the lead electrons.  The last term, $H_{\rm
tunneling}=\sum_{k\alpha\sigma}t_{\alpha}(c_{k\alpha\sigma}^{\dagger}d_{\sigma}
+H.c.)$, describes the tunneling of electrons from the leads onto and off the
dot with tunneling amplitudes $t_{\alpha}$. By using even and odd combinations
of lead electron states, the odd channel decouples, resulting in a
single-channel Anderson model with an effective tunnel matrix element $t$ given
by $t^2=t_{L}^2+t_{R}^2$. The model is then characterized by the lead-dot
tunneling rate $\Gamma= 2\pi N_{\rm F}t^2$, with $N_{\rm F}$ the lead electron
density of states at the Fermi level, the level position $\varepsilon_{0}$, and
the local Coulomb repulsion $U$. In the real system, the level position is
controlled by a gate voltage $V_{g}$ via $\varepsilon_0 = -|e|V_{g}+V_{\rm
offset}$, with $V_{\rm offset}$ an arbitrary offset voltage.  Equivalently,
varying the gate voltage $V_{g}$ controls the Fermi level $E_{\rm F}$ and thus
$\varepsilon_{0}-E_{\rm F}$. Without loss of generality, we shall set $E_{\rm
F}=0$ as our zero energy reference throughout. 

In describing the theoretical results, it is convenient to define a
dimensionless gate voltage ${\rm v}_g=(\varepsilon_0+U/2)/\Gamma$ which vanishes
at the particle-hole symmetric point of the Anderson model,
$\varepsilon_0=-U/2$, where the thermopower also (trivially) vanishes. Three
different physical regimes are relevant for characterizing and understanding the
linear thermoelectric transport properties of the model in the strongly
correlated case $U/\Gamma\gg 1$: (i) the charge fluctuation, or, mixed valence
regime, which corresponds to level positions  $\varepsilon_{0}$ lying within
$\Gamma/2$ of the Fermi level, i.e., $|\varepsilon_{0}|\leq \Gamma/2$. In this
regime, the charge on the dot, $n_{0}=\sum_{\sigma}\langle n_{d\sigma}\rangle$,
fluctuates  between approximately  $0$ and $1$ electrons so the average
occupation is approximately $n_{0}=1/2$ (with deviations from this value
depending on the precise value  of $\varepsilon_{0}$ in the above range).
Another mixed valence regime occurs when $|\varepsilon_{0}+U|\lesssim \Gamma/2$.
In this case, the charge on the dot fluctuates approximately between $1$ and $2$
electrons, and the average occupation of the dot is approximately $n_{0}=3/2$
(with the precise value depending on the location of  $\varepsilon_{0}+U$ in the
above range). For $\varepsilon_{0}>\Gamma/2$ or $\varepsilon_{0}+U < -\Gamma/2$,
the dot occupancy tends to zero, or two electrons, respectively, i.e., to an
even number of electrons on the dot. In this empty (or full) orbital regime,
many-body interactions become irrelevant, so that a description in terms of an
effective noninteracting model becomes possible. Finally, for $|\varepsilon_0 +
U/2| \lesssim U/2 -\Gamma/2$, i.e., between the above two mixed valence regimes,
the dot occupation approaches $1$, charge fluctuations are suppressed by the
Coulomb energy $U\gg \Gamma$, and low energy spin fluctuations give rise to the
Kondo effect.  For $U/\Gamma=8$, the Kondo regime occurs for dimensionless gate
voltages $-3.5 \lesssim  {\rm v}_{g}\lesssim +3.5$, the mixed valence regimes
occur for $-4.5 \leq {\rm v}_g\leq -3.5$ and $3.5 \leq {\rm v}_g \leq +4.5$ and
the empty (full) orbital regimes occur for ${\rm v}_{g}\geq 4.5$ and ${\rm
v}_{g}\leq -4.5$.   

%
\subsection{Linear thermoelectric transport calculations}
\label{subsec:linear-transport}
The linear thermoelectric transport properties of the above model are calculated in terms of the energy and temperature dependent spectral function of the dot, $A(E,T)$,  where $E$ is the excitation energy measured
relative to the Fermi level $E_{\rm F}$ and $T$ is the temperature\cite{Kim2002,Costi2010}. 

The conductance $G(T)$ and thermopower $S(T)$ can be written in terms of the zeroth, $I_0(T),$ and first, $I_1(T)$, moments of the spectral function $A(E,T)$ within the Fermi temperature window as follows:
\begin{align}
\frac{G(T)}{G_0}  &= I_{0}(T), \label{eq:gt}\\
S(T) & = -\frac{1}{|e|T}\frac{I_1(T)}{I_{0}(T)},\label{eq:st}
\end{align}
where  $G_0$ is the zero temperature conductance at midvalley, and the moments $I_{i=0,1}$ are given by
\begin{equation}
I_{i=0,1}(T) = \int_{-\infty}^{+\infty}-\frac{\partial f(E,T)}{\partial E} E^{i}\, A(E,T) dE,\label{eq:moments-def}
\end{equation}
where $f(E,T)$ is the Fermi distribution of the leads at temperature $T$
%

The above expression for $S(T)$ makes it clear that a sign change in the thermopower occurs due to a competition between the negative (electron-like) and positive (hole-like) enegy contributions of the
moment of the spectral function in the Fermi temperature window $-k_{\rm B}T\lesssim E \lesssim +k_{\rm B}T$. At particle-hole symmetric points, such as for integer fillings, e.g., exactly in the middle of even or odd 
Coulomb diamonds, the spectral function is symmetric about $E_{\rm F}$ and $S(T)$ vanishes identically. 

We use the numerical renormalization group (NRG) method\cite{Wilson1975,Bulla2008} to calculate $A(E,T)$. The NRG uses a discrete approximation to the continuum Anderson model and therefore results in a discrete
representation for the spectral function $A(E,T)$ as a set of delta function peaks at the excitation energies of the system, i.e., 
in the discrete Lehmann representation:
\begin{align}
A(E,T) = &\frac{1}{Z(T)}\sum_{m,n,\sigma}|\langle m|d_{\sigma}|n\rangle|^2(e^{-E_m/k_{\rm B}T}+e^{-E_n/k_{\rm B}T})\delta(E - (E_{n}-E_{m})),\label{eq:spectral-function}
\end{align}
where $E_m$ and $|m\rangle$ are the eigenvalues and eigenstates of the system and $Z(T)=\sum_{m}e^{-E_m/k_{\rm B}T}$ is
the partition function. For visualizing  $A(E,T)$, a broadening
of  the delta functions with Gaussians of widths proportional to the energies of the excitations $E=E_n-E_m$ is necessary. 
Since the NRG calculates the excitations on a logarithmic scale around $E_{\rm F}=0$, this broadening procedure inevitably leads to an overbroadening of the high-energy spectral features, such as the Hubbard peaks at $E=\varepsilon_0$ and $E=\varepsilon_0+U$, as compared to the low-energy features, such as the Kondo peak at $E_{\rm F}$, whose width is correctly resolved by this procedure. This overbroadening of the high energy features can be avoided, to a large extent, by evaluating the spectral function indirectly via the many-body self-energy\cite{Bulla1998}. However, the actual calculations of $G(T)$ and $S(T)$, from the moments $I_0$ and $I_1$, do not require a broadened spectral function: the integrations for $I_0(T)$ and $I_1(T)$ can be evaluated analytically using the discrete form of the spectral function to give
\begin{equation}
I_{i=0,1}(T) = \frac{1}{k_{\rm B}TZ(T)}\sum_{m,n,\sigma}(E_n-E_m)^{i}\frac{|\langle m|d_{\sigma}|n\rangle|^2}
{(e^{E_m/k_{\rm B}T}+e^{E_n/k_{\rm B}T})}.\label{eq:moments}
\end{equation}
In addition to avoiding any additional errors arising from a broadening procedure, this way of
calculating $I_0(T)$ and $I_1(T)$ can be carried out with high precision\cite{Yoshida2009,Merker2013}, since
the above expressions for these moments takes the same form as the calculation of thermodynamic observables
within the NRG. Such calculations are known to be essentially exact by comparisons with 
Bethe-Ansatz calculations\cite{Merker2012b}. Hence, the calculations for $S(T)$ and 
$G(T)$, using the moments $I_0(T)$ and $I_1(T)$ obtained with the discrete form of the spectral function, are also 
essentially exact at all temperatures and for all parameter values. In particular, as compared to a previous study \cite{Costi2010} which used the broadened spectral function to calculate $G(T)$ and $S(T)$, we are here able
to provide some additional insights concerning the detailed behavior of the thermopower (see next subsection).

\begin{figure}[H]
\centering 
\includegraphics[width=0.6\textwidth]{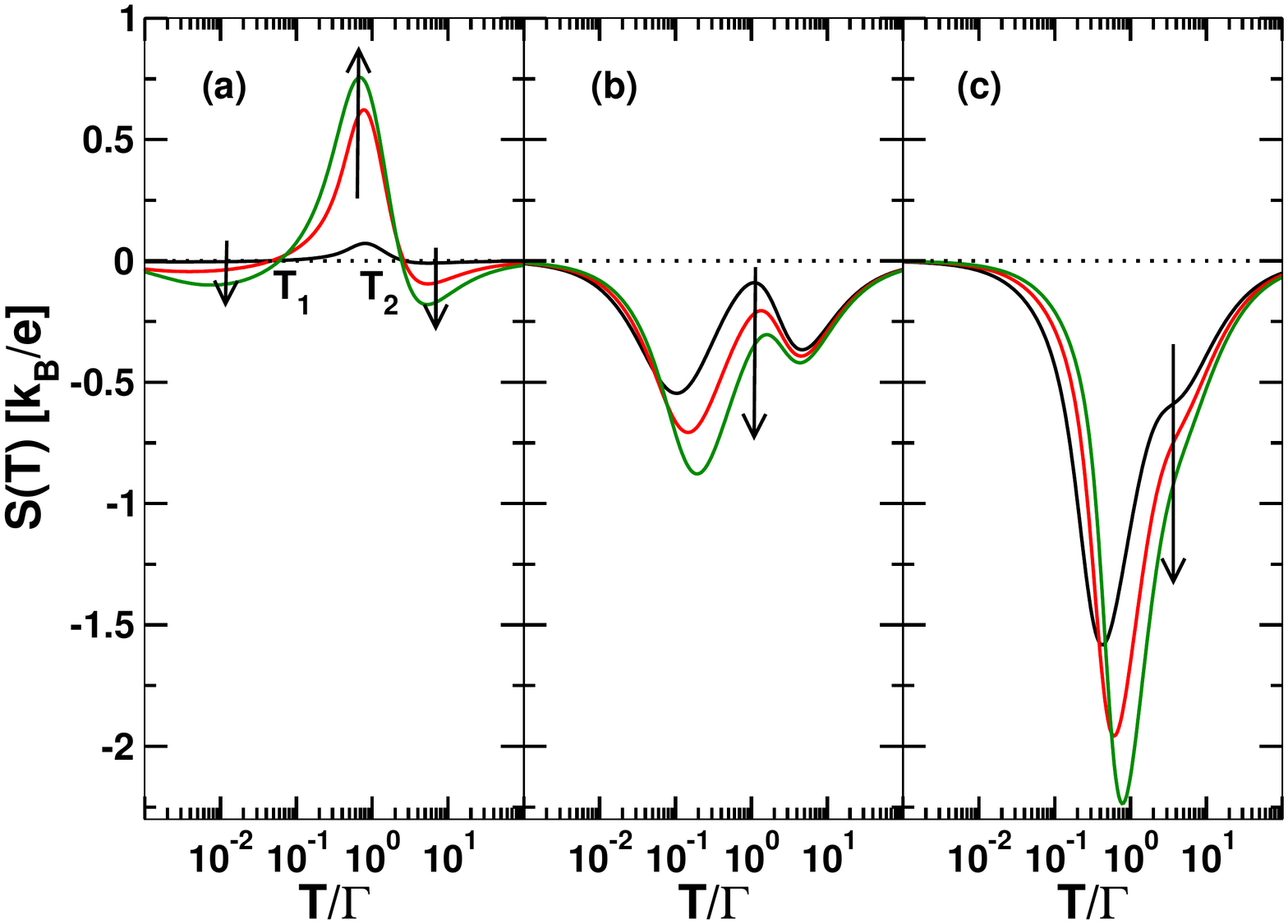}
\caption 
{
The thermopower $S(T)$ versus temperature $T/\Gamma$ for $u=U/\Gamma=8$ and different gate voltages ${\rm v}_g$. (a) Kondo regime: ${\rm v}_{g}=0.1,1.0$ and $1.8$. (b) Mixed valence regime: ${\rm v}_{g}=3.6, 3.8$ and $4.28$. (c) Empty orbital regime: ${\rm v}_{g}=5.1, 6.0$ and $6.9$. Vertical arrows indicate the trends with increasing ${\rm v}_g$. 
The gate voltage dependent zeros of $S(T)$ at $T=T_{1}$ and $T=T_{2}$ are also indicated. 
}
\label{fig:ST}
\end{figure}
\subsection{Characterization of the results for $S_{{\rm v}_g}(T)$ versus $T$ at different ${\rm v}_g$}
\label{subsec:characterization-theory}
The behaviour of the thermopower as function of temperature exhibits different characteristic
behaviour in the three different regimes of the model \cite{Costi2010}. It suffices to consider 
just ${\rm v}_g>0$, since the thermoelectric properties for ${\rm v}_g<0$ can be obtained
via a particle-hole transformation: $G_{{\rm v}_g}(T) = G_{{\rm
    -v}_g}(T)$ and $S_{{\rm v}_g}(T)=-S_{{\rm -v}_g}(T)$. To
illustrate the trends, we have used $u=8$, however, for comparisons
with the measurements, calculations were also performed at the
experimentally determined value of $u=22$.

For gate voltages ${\rm v}_g$ in the mixed valence and empty (full) orbital regimes, the thermopower $S_{{\rm v}_g}(T)$ exhibits no sign changes as function of $T$, see Fig.~\ref{fig:ST}(b)-\ref{fig:ST}(c). The mixed valence regime does, however, exhibit two peaks in $S_{{\rm v}_g}(T)$, which merge into a single peak (and a remnant shoulder) in the empty (full) orbital regimes [see Fig.~\ref{fig:ST}(c)]. 

The most interesting behaviour in the temperature dependence of the thermopower is in the Kondo regime of gate voltages. For fixed ${\rm v}_g$, $S_{{\rm v}_g}(T)$ exhibits two sign changes upon increasing $T$ from $T\ll T_{\rm K}({\rm v}_g)$ to $T\gg \Gamma$ [see Fig.~\ref{fig:ST}(a)]: one at $T=T_{1}({\rm v}_g)\gg T_{\rm K}({\rm v}_g)$ and a second one at $T=T_2({\rm v}_g)\gtrsim \Gamma\gg T_{\rm K}({\rm v}_g)$, where $T_{\rm K}({\rm v}_g)$ is the gate voltage dependent Kondo temperature defined by
\begin{align}
\frac{T_{\rm K}({\rm v}_g)}{\Gamma} = & \sqrt{\frac{U}{4\Gamma}}e^{-\pi |\varepsilon_{0}||\varepsilon_{0}+U|/\Gamma U}
\equiv \sqrt{\frac{u}{4}}e^{-\pi(u^2/4-{\rm v}_{g}^2)/u},\label{eq:tk-vs-vg}
\end{align}
and we have introduced the dimensionless Coulomb energy $u=U/\Gamma$. 
The gate voltage dependence of $T_{1}$ and $T_2$ have been described previously \cite{Costi2010}. We summarize and expand on this here.  $T_{1}({\rm v}_g)$ and $T_{2}({\rm v}_g)$ increase (decrease) monotonically from their minimum (maximum) values at midvalley (${\rm v}_g=0$) with increasing ${\rm v}_g$ and merge to a common value at a critical gate voltage
${\rm v}_{g}={\rm v}_g^c = -\frac{1}{2}(1-u)$, where ${\rm v}_g^{c}$ is the gate voltage for entry into the mixed valence regime [the mixed valence regime at ${\rm v}_{g}<0$ has ${\rm v}_g^c = +\frac{1}{2}(1-u)$].  
The midvalley values $T_{1}$ and $T_{2}$ are denoted by $T_{1}^{\rm min}=T_{1}({\rm v}_g\to 0)$ and $T_{2}^{\rm max}=T_{2}({\rm v}_g\to 0)$, respectively, and are shown in Table~\ref{Table1}  for a range of $u=U/\Gamma$ together with the corresponding midvalley Kondo scales $T_{\rm K}({\rm v}_g\to 0)$.
We note that $T_{1}^{\rm min}$ and $T_{2}^{\rm max}$ exist as limit values and are finite, despite the fact that
the thermopower vanishes identically at midvalley: $S(T)_{{\rm v}_g=0}\equiv 0$.  $T_{1}({\rm v}_g)$ and $T_{2}({\rm v}_g)$ are approximate crossing points (isosbectic points) and are
a common feature of strongly correlated systems \cite{Vollhardt1997}. In general, as we verify in more detail below, neither $T_{1}({\rm v}_g)$ nor $T_{2}({\rm v}_g)$ are low energy scales, but the former can approach several $T_{\rm K}$ for smaller $U/\Gamma$ or for values of ${\rm v}_g$ approaching the mixed valence regime when both  $T_{1}({\rm v}_g)$ and the Kondo scale $T_{\rm K}({\rm v}_g)$ become larger and can be of comparable magnitude. 

From Table~\ref{Table1} we see that $T_{2}({\rm v}_g\to 0)\gtrsim \Gamma\gg T_{\rm K}$, i.e., a high energy scale of the Anderson model. A detailed analysis of the data in Table~\ref{Table1} shows that for $u\gg 1$,
\begin{equation}
T_{2}({\rm v}_g\to 0)/\Gamma =-0.466 + 0.394u. \label{eq:t2-vs-u}
\end{equation}
At $T\approx T_{2}({\rm v}_g\to 0)$ we also find a minimum in the dot occupation $n_{0}(T)$. This minimum is absent in the other regimes. For fixed $u$, we find that $T_2({\rm v}_g)$ versus ${\rm v}_g$ correlates with the dot level position $\varepsilon_{0}$.  
\begin{table}[t]
\begin{tabular}{cccc}
\hline
\hline
\multicolumn{1}{c}{$u$} & 
\multicolumn{1}{c}{$T_{1}^{\rm min}/\Gamma$} & 
\multicolumn{1}{c}{$T_{2}^{\rm max}/\Gamma$} & 
\multicolumn{1}{c}{$T_{\rm K}({\rm v}_g=0)/\Gamma$}\\
\hline
$4$   & $0.0983$ & $1.1009$ & $4.321\times 10^{-2}$\\
$6$   & $0.0612$ & $1.9021$ & $1.100\times 10^{-2}$\\
$8$   & $0.0426$ & $2.6924$ & $2.641\times 10^{-3}$\\
$10$ & $0.0322$ & $3.4800$ & $6.138\times 10^{-4}$\\ 
$12$ & $0.0258$ & $4.2666$ & $1.398\times 10^{-4}$\\ 
$14$ & $0.0215$ & $5.0480$ & $3.138\times 10^{-5}$\\ 
$16$ & $0.0185$ & $5.8366$ & $6.974\times 10^{-6}$\\ 
$22$ & $0.0131$ & $8.1969$ & $7.347\times 10^{-8}$\\ 
\hline
\hline
\end{tabular}
\caption
{$T_{1}^{\rm min}/\Gamma=T_{1}({\rm v}_g/\Gamma\to 0)$ and $T^{\rm max}_2/\Gamma=T_{2}({\rm v}_g/\Gamma\to 0)$
for different values of $u=U/\Gamma$ together with the midvalley 
Kondo scale $T_{\rm K}({\rm v}_g=0)/\Gamma$, 
}
\label{Table1}
\end{table}

\begin{figure}[H]
\centering 
\includegraphics[width=0.6\textwidth]{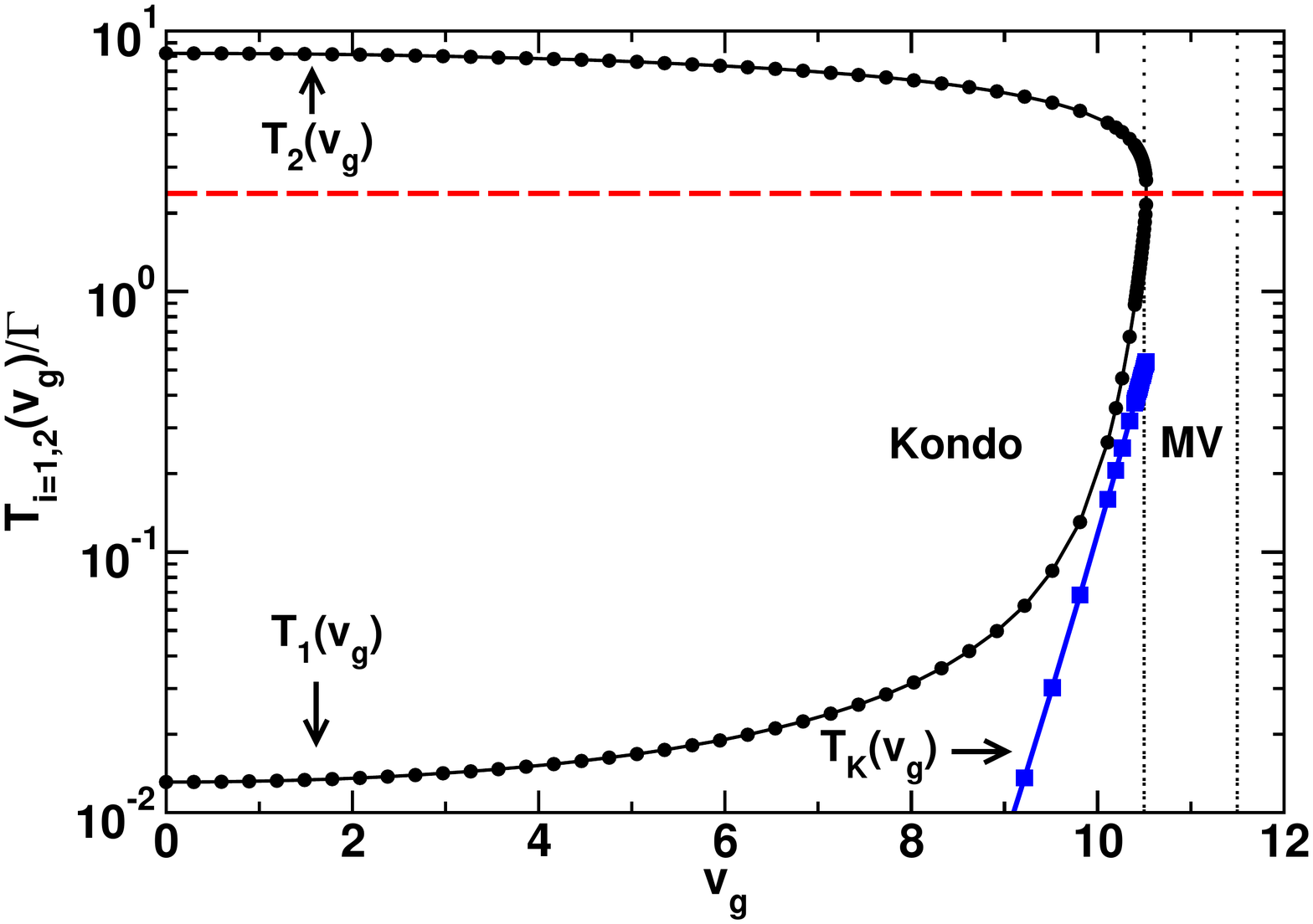}
\caption 
{
$T_{1}$ and $T_{2}$ (in units of $\Gamma$) versus dimensionless gate voltage ${\rm v}_{g}$ for $u=22$ of the experiment. 
$T_{1}$ and $T_{2}$ attain their minimum and maximum values, respectively, at midvalley (${\rm v}_g=0$) and they
merge upon leaving the Kondo regime and entering the mixed valence (MV) regime at ${\rm v}_{g}^c\approx -1/2 +u/2=10.5$ (corresponding to $\varepsilon_{0}=-\Gamma/2$), and take the value $T_{1}({\rm v}_{g}^c)/\Gamma=T_{2}({\rm v}_{g}^c)/\Gamma\approx 2.38$ 
as indicated by the red horizontal dashed line. The vertical dotted lines delimit the mixed valence regime from the Kondo regime to the left (${\rm v}_g\leq 10.5$) and the empty
orbital regime to the right (${\rm v}_g\geq 11.5$). The Kondo scale $T_{\rm K}({\rm v}_g)$ versus ${\rm v}_g$ [from Eq.~(\ref{eq:tk-vs-vg})] is also shown (blue solid line with square symbols). The exponential decrease of  $T_{\rm K}({\rm v}_g)$ on approaching midvalley is not visible 
on the scale of the plot. Generally, $T_{\rm K}({\rm v}_g)\ll T_{1}({\rm v}_g)$, except on approaching the mixed valence regime when 
$T_{1}({\rm v}_g)$ and $T_{\rm K}({\rm v}_g)$ can become comparable, as seen in the figure.
}
\label{fig:t12}
\end{figure}

Turning now to $T_{1}$, an analysis of the data in  Table~\ref{Table1} for $T_{1}({\rm v}_g\to 0)$ gives
\begin{equation}
T_{1}({\rm v}_g\to 0)/\Gamma = c_{0} u^{-{\alpha}},\label{eq:t1min-vs-u}
\end{equation}
with  $c_{0} \approx 0.53$ and $\alpha\approx 1.2$. Clearly, unlike $T_{\rm K}({\rm v}_g)$, $T_{1}({\rm v}_g\to 0)$ is not exponentially
decreasing with increasing $u$ and cannot, therefore, be considered a low energy scale\footnote{This result was anticipated but not demonstrated in Ref.~\cite{Costi2010}}. A low energy scale characterizing the Kondo induced peak in the low temperature thermopower is the position $T_{p}({\rm v}_g)$ of this
peak, which has been shown to scale with $T_{\rm K}({\rm v}_g)$ \cite{Costi2010}. One sees from Table~\ref{Table1} that typical
values of $T_{1}$ at midvalley are $1-10\%$ of $\Gamma$, even for large values of $u$. While these midvalley values might appear to be 
irrelevant for experiments, as $S(T)$ vanishes identically there, they are nevertheless relevant since they can be taken as
lower bounds for $T_{1}$ at finite gate voltages, where the thermopower $S(T)$ is finite. Moreover, the gate voltage dependence of
 $T_1$ is rather weak in the Kondo regime and only increases rapidly on approaching the mixed valence regime, where
$T_1$ approaches a value of $O(\Gamma)$, eventually merging with $T_2$ at  ${\rm v}_g= {\rm v}_g^c$ as shown in Fig.~\ref{fig:t12} for the case $u=22$ relevant to the experiment.  This allows the Kondo induced sign changes in the thermopower at low, but not exponentially low temperature scales, to be accessed in the temperature range $T_2\gtrsim T\gtrsim T_{1}$. One sees, for example, that for the whole range $-U/2\lesssim \varepsilon_0/\Gamma\lesssim -1$ (i.e., $0\leq {\rm v}_g \leq -1+u/2=10.5$) in the Kondo regime that $T_1/\Gamma$ ranges in value between $0.01$ and $0.2$. Thereby, the experiment is able to access the temperature range (i) for $T<T_1$ where $S_T({\rm v}_g>0)<0$ as well as the temperature range (ii) for $T_1 < T < T_2$ where $S_T({\rm v}_g)$ can be positive or negative depending on ${\rm v}_g$. 
From Fig.~\ref{fig:t12}, one can read off, for any fixed gate voltage in the Kondo regime, the temperature $T_1=T_{1}({\rm v}_g)$ at which the thermopower changes sign upon increasing temperature through $T_1$, or, for any fixed temperature $T$, the gate voltage ${\rm v}_g^{0}$ at which 
$S_T({\rm v}_g$ versus ${\rm v}_{g}$ changes sign (see next section).
\begin{figure}[H]
  \centering 
  \includegraphics[width=0.6\linewidth]{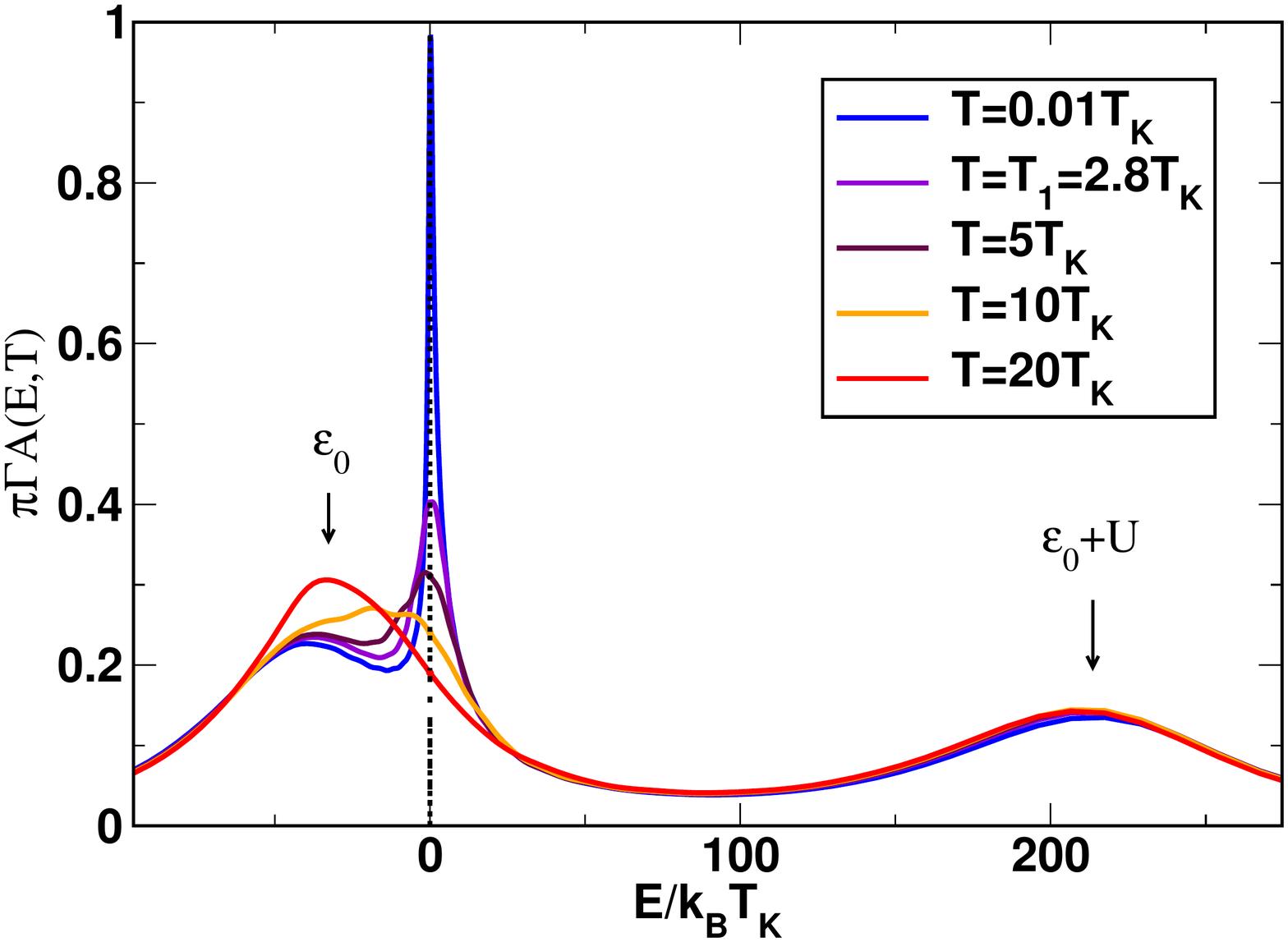}
  \caption{
    \label{fig:spectra}
Normalized spectral function $\pi\Gamma A(E,T)$ versus excitation energy $E$ (in units of $k_{\rm B}T_{\rm K}$) 
at several temperatures, illustrating the 
destruction of the Kondo resonance with temperature and how its spectral weight gets redistributed 
to higher energies. Parameters: $u=U/\Gamma=8$ and $\varepsilon_0=-1.5\Gamma$.
}
\end{figure}

A microscopic understanding of the sign changes in the thermopower at $T_1$ and $T_2$ requires a closer look at the
temperature dependence of the spectral function $A(E,T)$ for fixed gate voltage (local level position) in the Kondo regime.
This is shown in Fig.~\ref{fig:spectra} 
for $u=8$ and $\varepsilon_{0}=-1.5\Gamma$ (${\rm v}_g= 2.5$) at several temperatures. 
At temperatures $T\ll T_{\rm K}<T_{1}({\rm v}_g)$ we have a fully developed Kondo resonance pinned close to, but
slightly above the Fermi level, thereby yielding a negative thermopower due to more hole-like excitations ($E>0$)
as compared to electron-like  ($E<0$) excitations in the Fermi temperature window $|E|\lesssim k_{\rm B}T$. Alternatively, one also sees this from a low temperature Sommerfeld expansion (which, however, is only valid for $T\ll T_{\rm K}$):
\begin{equation}
S(T) = -\frac{k_{\rm B}}{|e|}\frac{\pi^2}{3}k_{\rm B}T\frac{1}{A(0,T)}\frac{\partial A}{\partial E}|_{E=0}.\label{eq:sommerfeld}
\end{equation}
Upon increasing temperature, the Kondo resonance is gradually destroyed on a temperature scale of order $10T_{\rm K}$.
In the process, spectral
weight from the Kondo resonance is  redistributed to higher energies $|E|>k_{\rm B}T_{\rm K}$. 
Initially, most of this weight (for this particular gate voltage ${\rm v}_g>0$) is shifted to below, rather than above, 
the Fermi level. Consequently the contribution of electron-like
excitations in $I_{1}(T)$ increase relative to that of the hole-like
excitations and this eventually results in a sign change of $I_{1}(T)$
and hence of $S(T)$ at $T=T_{1}$. The above picture, holds generally,
and explains the sign change of the thermopower at $T=T_{1}$ in terms of
the temperature dependent spectral weight shift of the asymmetrically
located Kondo resonance.
A further increase in temperature expands the Fermi window sufficiently such that the contributions of the incoherent (Hubbard) satellite peaks at $E=\varepsilon_{0}<0$ and $E=\varepsilon_{0}+U > 0$ become determining factors in the sign of $I_{1}(T)$. The weight of these
excitations is approximately $2-n_{0}(T)$ and $n_{0}(T)$, respectively. The second sign change at $T_2$ corresponds to an increasing population of the hole-like excitations at $E=\varepsilon_{0}+U $ and a decreasing population of the electron-like excitations at $E=\varepsilon_{0}$, i.e., at the minimum of $n_{0}(T)$ vs $T$. Such a minimum, responsible for the sign change of $I_{1}(T)$,
at $T=T_2$, and hence of $S(T)$, is indeed only observed in the Kondo regime\cite{Costi2010}.

\subsection{Characterization of $S_{T}({\rm v}_g)$ versus ${\rm v}_g$ at different $T$}
 
\begin{figure}[H]
    \centering 
  \includegraphics[width=0.69\linewidth]{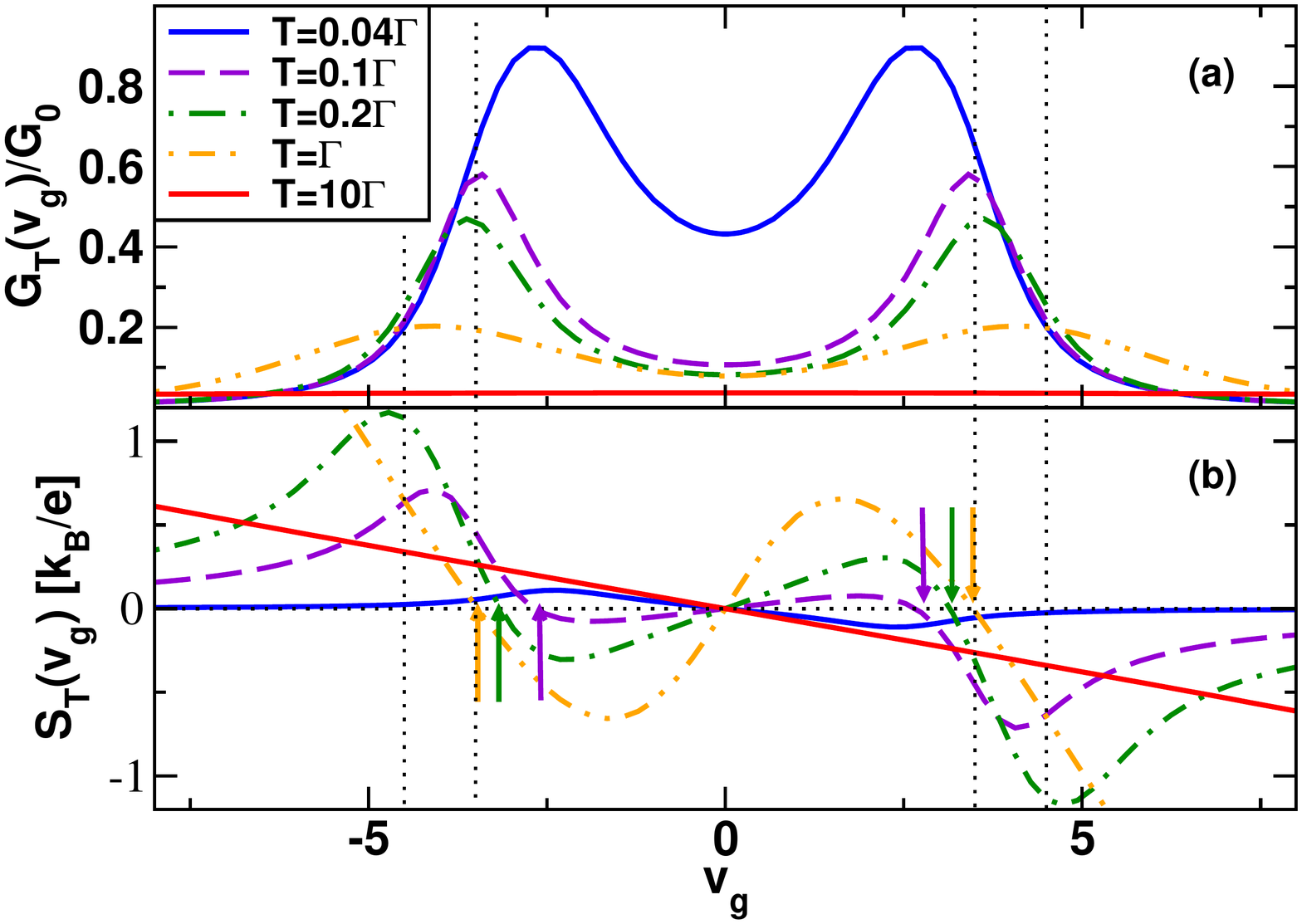}
   \caption{Gate voltage dependence of, (a), the conductance, $G_T({\rm v}_g)$, and, (b), the thermopower, $S_{T}({\rm v}_g)$, at different temperatures and for $u=U/\Gamma=8$. Vertical dotted lines delineate the central Kondo regime ($|{\rm v}_g|\lesssim 3.5$) from the mixed valence ($3.5\lesssim |{\rm v}_g|\lesssim 4.5$) and empty (full) orbital regimes ($|{\rm v}_{g}|\gtrsim 4.5$).  
At $T=0.04\Gamma<T_{1}^{\rm min}=0.042\Gamma$, the thermopower does not change sign at any gate voltage  
(except trivially at ${\rm v}_g=0$). Similarly, at $T=10\Gamma>T_{2}^{\rm max}=2.69\Gamma$, there is no sign change  
at finite ${\rm v}_g$. In the Kondo regime, and for $T_{1}({\rm v}_g)<T<T_{2}({\rm v}_g)$, a characteristic additional sign change  
of the thermopower at finite gate voltage is found (vertical arrows), e.g., for $T=0.1\Gamma, 0.2\Gamma$ and $T=\Gamma$. 
}
  \label{fig:Sgate}
\end{figure}

The above characteristic temperature dependence of the thermopower in the different regimes of gate voltage can be straightforwardly translated into different characteristic gate voltage dependencies  of the thermopower, $S_{T}({\rm v}_g)$, in different temperature ranges: (i) $T\leq T_{1}^{\rm min}$, (ii), $T_{1}^{\rm min}< T < T_{2}^{\rm max}$, and, (iii), $T>T_{2}^{\rm max}$. 
This gate voltage dependence, for temperatures in each of the above ranges, is shown in Fig.~\ref{fig:Sgate}(b) 
together with that of the conductance [Fig.~\ref{fig:Sgate}(a)].
In the first and last temperature ranges (see also Fig.~\ref{fig:ST}), the thermopower exhibits no sign change as a function of ${\rm v}_g$ (except the trivial one at ${\rm v}_g=0$), as depicted in Fig.~\ref{fig:Sgate}(b) for the $T=0.04\Gamma$ and $T=10\Gamma$ curves. In the temperature range (ii) $S_T({\rm v}_g)$ has the opposite sign to that in (i) and (iii) in a range of gate voltages about ${\rm v}_g=0$ [delimited by the vertical arrows in Fig.~\ref{fig:Sgate}(b)]. Outside  this range, the thermopower, for fixed ${\rm v}_g$, is of the same sign at all temperatures [see the $T=0.1\Gamma,0.2\Gamma$ and $T=\Gamma$ curves in Fig.~\ref{fig:Sgate}(b)]. The change in sign of $S_{T}({\rm v}_g)$  versus ${\rm v}_{g}$ at the boundary between these ranges [see the vertical arrows in  Fig.~\ref{fig:Sgate}(b)] and for
temperatures in the range (ii) is characteristic  of the thermopower of a Kondo correlated quantum dot. It is absent for
weakly correlated quantum dots ($u\lesssim 1$) where there is no sign change in either the $T$- or ${\rm v}_g$-dependence
of the thermopower (except the trivial one at ${\rm v}_g=0$) \cite{Costi2010}.
\label{subsec:gate-voltage-dependence}

\end{document}